\documentclass[review]{elsarticle}

\usepackage{amsmath,epsfig,amssymb,multirow,graphics,graphicx}
\usepackage{color}
\usepackage{dcolumn}    
\usepackage{bm}
\usepackage{subfigure}
\usepackage{amsfonts}
\usepackage{mathrsfs}
\usepackage{fancyhdr}
\usepackage{indentfirst}
\usepackage{chngpage}
\usepackage{color}
\usepackage{url}
\usepackage{multirow}
\usepackage{diagbox}
\usepackage{algorithm,algpseudocode,float}
\usepackage{setspace}
\usepackage{breakurl}
\usepackage{hyperref}
\usepackage{geometry}
\usepackage{enumerate}
\usepackage{epstopdf}
\usepackage{color}

\usepackage{booktabs, multirow} 
\usepackage{soul}
\usepackage[table]{xcolor} 
\usepackage{threeparttable}
\biboptions{sort&compress}

\makeatletter
\newenvironment{breakablealgorithm}
{
	\begin{center}
		\refstepcounter{algorithm}
		\hrule height.8pt depth0pt \kern2pt
		\renewcommand{\caption}[2][\relax]{
			{\raggedright\textbf{\ALG@name~\thealgorithm} ##2\par}%
			\ifx\relax##1\relax 
			\addcontentsline{loa}{algorithm}{\protect\numberline{\thealgorithm}##2}%
			\else 
			\addcontentsline{loa}{algorithm}{\protect\numberline{\thealgorithm}##1}%
			\fi
			\kern2pt\hrule\kern2pt
		}
	}{
		\kern2pt\hrule\relax
	\end{center}
}

\newcommand\ASTART{\bigskip\noindent\begin{minipage}[b]{0.5\linewidth}}

\newcommand\AENDSKIP{\end{minipage}\bigskip}
\newcommand\AEND{\end{minipage}}

\journal{Journal of \LaTeX\ Templates}

\begin{document}

\begin{frontmatter}

\title{Associative Learning Mechanism for Drug-Target Interaction Prediction}
\tnotetext[mytitlenote]{Fully documented templates are available in the elsarticle package on \href{http://www.ctan.org/tex-archive/macros/latex/contrib/elsarticle}{CTAN}.}

\author[mymainaddress]{Zhiqin Zhu}
\ead{zhuzq@cqupt.edu.cn}

\author[mymainaddress]{Zheng Yao}
\ead{Lamouryz2019@gmail.com}

\author[mysecondaryaddress]{Guanqiu Qi\corref{mycorrespondingauthor}}
\cortext[mycorrespondingauthor]{Corresponding author}
\ead{qig@buffalostate.edu}

\author[mysecondaryaddress]{Neal Mazur}
\ead{mazurnm@buffalostate.edu}

\author[mythirdaddress]{Baisen Cong\corref{mycorrespondingauthor}}
\cortext[mycorrespondingauthor]{Corresponding author}
\ead{bcong@dhdiagnostics.com}

\address[mymainaddress]{College of Automation, Chongqing University of Posts and Telecommunications, Chongqing, 400065, China}
\address[mysecondaryaddress]{Computer Information Systems Department, State University of New York at Buffalo State, Buffalo, NY, 14222, USA}
\address[mythirdaddress]{Data Scientist, Diagnostics Digital DH (Shanghai) Diagnostics Co.,Ltd., a Danaher company}

\begin{abstract}

As a necessary process in drug development, finding a drug compound that can selectively bind to a specific protein is highly challenging and costly.
Drug-target affinity (DTA), which represents the strength of drug-target interaction (DTI), has played an important role in the DTI prediction task over the past decade.
Although deep learning has been applied to DTA-related research, existing solutions ignore fundamental correlations between molecular substructures in molecular representation learning of drug compound molecules/protein targets.
Moreover, traditional methods lack the interpretability of the DTA prediction process.
This results in missing feature information of intermolecular interactions, thereby affecting prediction performance.
Therefore, this paper proposes a DTA prediction method with interactive learning and an autoencoder mechanism.
The proposed model enhances the corresponding ability to capture the feature information of a single molecular sequence by the drug/protein molecular representation learning module and supplements the information interaction between molecular sequence pairs by the interactive information learning module.
The DTA value prediction module fuses the drug-target pair interaction information to output the predicted value of DTA. Additionally, this paper theoretically proves that the proposed method maximizes evidence lower bound (ELBO) for the joint distribution of the DTA prediction model, which enhances the consistency of the probability distribution between the actual value and the predicted value.
The experimental results confirm mutual transformer-drug target affinity (MT-DTA) achieves better performance than other comparative methods.

\end{abstract}

\begin{keyword}
\texttt{}binding affinity, drug-target interaction, transformer
\MSC[2020] 00-01\sep  99-00
\end{keyword}

\end{frontmatter}

\section{Introduction}\label{sec1}

In the process of drug discovery, it is costly to apply high-throughput screening methods to determine the affinity between target proteins and drugs\cite{jade2021virtual, jarada2021snf, ding2020identification}.
As massive biomedical datasets are collected and open to public, they are used to make computational predictions of possible active drugs, thereby improving the efficiency of drug discovery \cite{1klebe2006virtual, gupta2021artificial, sabe2021current}.
Drug-target interactions (DTI) can reflect the tightness of drug-target binding, so DTI prediction has gradually become an important research direction for drug screening and finding new drugs \cite{2cheng2017large, soh2022hidti, yadav2022role}.
Owing to the wide application of artificial intelligence technology in various fields, machine learning-based models have shown excellent performance in predicting DTI.
Existing models typically accomplish DTI prediction tasks by modeling drug-target pair interactions.
In general, machine learning-based models can be divided into two categories based on how they work,
structure-based and drug-target affinity (DTA)-based models.
Structure-based machine learning models \cite{4cao2014computational, 5ozturk2016comparative} commonly ues binary classification methods to predict DTI.
The affinity relation or similarity between the SMILES character sequence of the drug molecule and the protein molecule sequence is used to determine whether the drug interacts with the target.

As the main disadvantage of such methods, the binarization operation does not consider the feature information between the drug-target pair,
and simply divides it into interactive and non-interactive categories.
Additionally, the prediction performance is limited by the division of the classification threshold.
DTA-based models \cite{6gomez2018automatic, 7wen2017deep} commonly use regression tasks to predict DTA.
DTA value as an important quantitative indicator of drug-target interaction relation can be used to predict the binding strength of drug-target molecules in drug screening and DTI process.
The value of DTA is usually measured by ligand binding assays Kd, inhibition constant Ki, or half maximal inhibitory concentration IC50 \cite{8pahikkala2015toward, kao2021toward}.
The DTI prediction between drug molecules/protein target molecules and protein target molecules/drug molecules is completed in this collaborative process by establishing the characteristic relations between the above measurement information and the known molecular structure information in the existing biological database.
Combined with known DTA information, the DTI prediction between drug molecules/protein target molecules and protein target molecules/drug molecules is completed.
Such methods do not need to include information on the similarity between drugs and between targets, and the continuous value DTA contains more information about drug-target pair interactions.

The DTA value can also be combined with some biochemical information-rich heterogeneous data to further improve the accuracy of DTI prediction, such as drug-disease association information, protein-protein interaction information, etc. \cite{wan2019neodti}
Therefore, they have higher training efficiency and stronger expansion.
Due to this advantage, machine learning methods based on DTA prediction are widely used in DTI prediction tasks.
Early-stage models, such as KronRLS \cite{8pahikkala2015toward}, utilized a similarity measure matrix of drug-target pairs to infer drug-target pair interactions.
But only linear dependencies in the training data are captured, and highly correlated features may lead to data redundancy.
In this regard, DeepDTA \cite{10ozturk2018deepdta} introduced a convolutional neural network (CNN) model to predict drug-target pair interactions by extracting molecular sequence features of drug-target pairs through CNN.
DeepConv-DTI \cite{11lee2019deepconv} shows that latent representations of the subsequences of a molecular sequence can be used to predict DTI.
Considering that the molecular structure may be more in line with the biochemical relation of drug-target pair interactions, GraphDTA \cite{12nguyen2021graphdta} introduced graph-based models.
Drug SMILES character sequences and protein molecular sequences are converted into molecular structures, and the molecular structure forms of drugs and proteins are used as model inputs.
Although the above models perform well in the drug-target interaction prediction, two main challenges remain to be investigated.

\begin{itemize}
\item
The interdependency between the SMILES character sequence of the drug and the sequence of the protein molecule is not considered, which means that the model lacks the global connection within the sequence.
\item
Introducing structural information into the model adds significant time and resource costs.
\end{itemize}

Additionally, protein 3D structure is only the prediction result of molecular structure, and the use of prediction information leads to an increase in model error rate.
Therefore, it is important to find a suitable method to establish the binding relations between drug molecules and protein molecules.
This approach can help machine learning models better capture the characteristic information.
Recently, transformer \cite{13kalakoti2022transdti} was introduced into various models \cite{14huang2021moltrans, zhang2022deepmgt}, and the corresponding results confirmed that it performed well in extracting the feature information of drugs and protein sequences.
Such models link global feature information within a single molecular sequence, but ignore the mechanism of interactions between pairs of molecular sequences.
Molecules are sub-structural \cite{15karlova2021molecular}, and the feature information should contain the overall connection of intra-molecular sub-structural interactions and atomic interactions.

The substructure of drug molecules and protein molecules is shown in Fig. 1, and the node in a red area represents a substructure.
The corresponding expression of the molecular substructure in the molecular sequence is listed above the molecular structure.
The value of the DTA used to represent the strength of the drug-target interaction is measured continuously \cite{16tanoori2021drug}. Specifically, continuously measured DTA values are derived from intermolecular non-covalent interactions, which contain interconnections between a large number of molecular substructures.
Therefore, the connections between substructures within a single molecule and the interactions between substructures between pairs of molecules cannot be ignored \cite{moon2022pignet}.

\begin{figure}[ht]
	\centering
    \includegraphics[height=4.5cm]{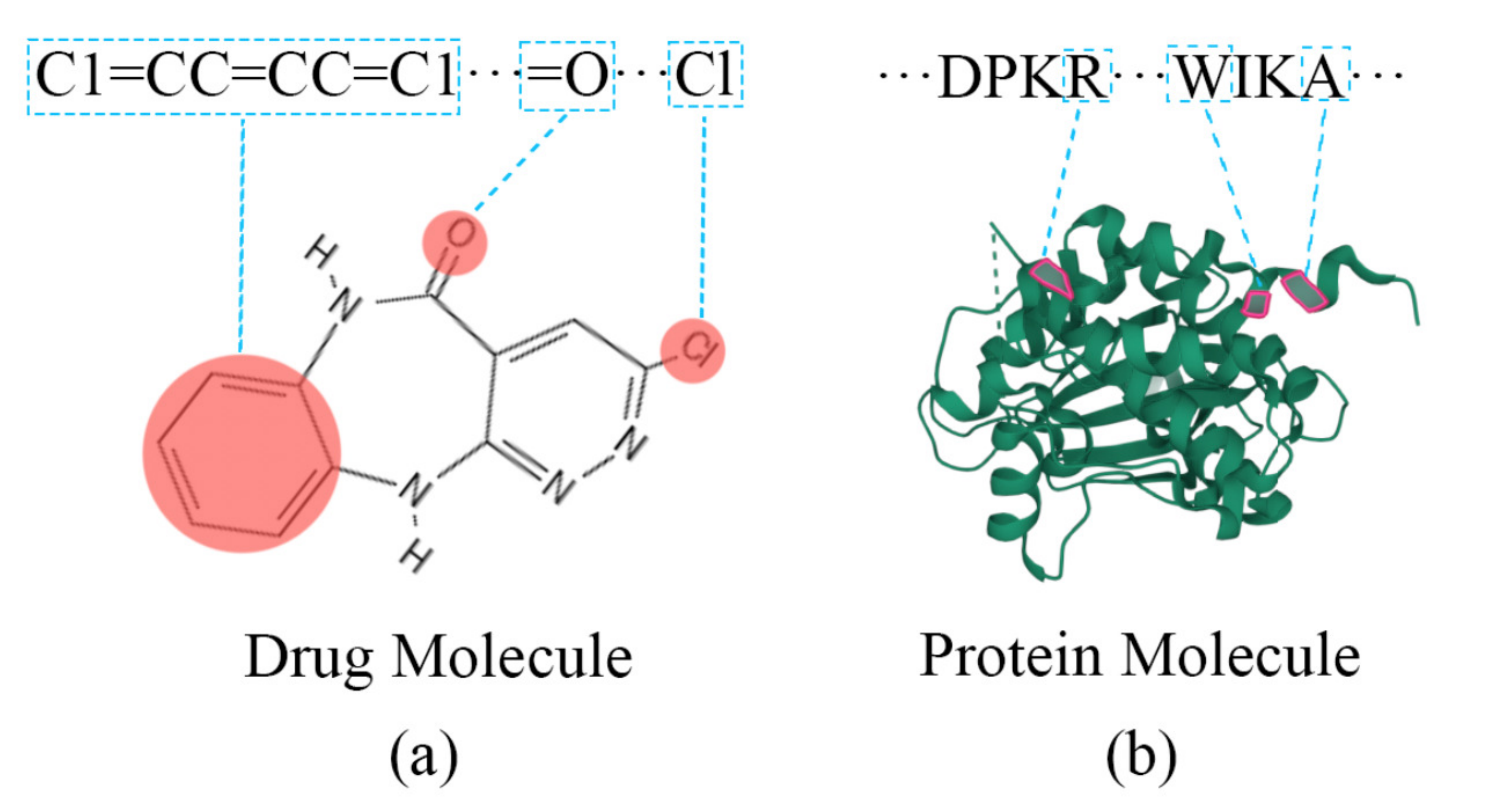}
	\caption{The substructures of drug molecules shown in (a) and protein molecules shown in (b). The red area represents the intramolecular substructure, and the blue dashed frame represents the molecular sequence expression corresponding to the substructure.}
	\label{fig1}
\end{figure}

In order to solve the shortcomings of the above-mentioned methods, this paper proposes an autoencoder model of mutual transformer-drug target affinity (MT-DTA) to predict DTA.
The SMILES character sequences and protein sequences of drugs are the inputs of the proposed model.
First, considering the correlation between substructures and single atoms in molecular sequences, this paper proposes a drug/protein molecular representation learning module to model the input SMILES character sequences and protein molecular sequences.
The ability of the proposed method to capture the feature information of a single molecule sequence is enhanced through this module.
SMILES character sequences and protein molecule sequences are converted into character sequences represented by integer codes using a dictionary.
In this way, the context connection between sequences can be established more conveniently and quickly \cite{MANKU2022108453}.
And substructures can be expressed as combinations of different atoms, as shown in Fig. 1.

The converted SMILES character sequence has 64 different characters for drug molecule sequences and 25 different characters for protein molecule sequences.
The convolutional neural networks (CNN) block is used to extract the basic feature information of the input sequence and establish the adjacent relations between substructures and atoms.
The transformer block is used to establish interactive information links between atoms and substructures in the input sequence\cite{17vaswani2017attention, le2021transformer}.
The correlation between atoms not only describes the relative position information, but also enhances the diversity learning of atoms.
The interactive information learning module complements the information interaction between molecular sequence pairs.
Information bridges are built for atoms and substructures within two independent molecular sequences to simulate drug-target interactions.
Feature information is represented by latent feature variables in the autoencoder model.
The drug target affinity value prediction module fuses the drug-target pair interaction information, and finally outputs the DTA prediction value.
Its loss is defined by the affinity error between the true and predicted DTA values.

In particular, the corresponding drug/protein molecular representation learning module builds two structurally similar decoder modules to reconstruct drug structures or target sequences.
The proposed model enhances the lateral connections between substructures and atoms in a single molecular sequence through the drug/protein molecular representation learning module, thereby making the feature representation more comprehensive and sufficient.
The learning path of substructure and atomic interaction information between molecular sequences is supplemented by the interaction information learning module.
In addition, using two decoders to reconstruct the drug structure or target sequence, the model can also generate new drugs with similar targets to the input drugs or new target sequences with similar structure and function to the input target sequences.
This paper has three main contributions as follows.

\begin{itemize}
\item
Aiming at the feature information of the correlation between substructures and atoms in drug molecular sequences and protein molecular sequences, a drug/protein molecular representation learning module is proposed.
Both local and global information links between substructures and atoms within a single molecular sequence are supplemented.
\item
An interactive learning mechanism is proposed for intramolecular and intermolecular molecular substructure interactions and affinity measurements.
Through the self-attention, the interaction relation between the target molecular sequence and the ligand molecular sequence and the interaction relation between the latent feature variables are established, which promotes the spatial connection between molecular sequences.
\item
Assume that the posterior distribution of the input data follows a normal distribution \cite{20li2021co}.
The proposed method theoretically proves that MT-DTA is the maximization of the evidence lower bound (ELBO) for the joint distribution of the affinity prediction model.
It makes the distribution of the actual value of DTA and the corresponding predicted value maintain a good consistency, which ensures the validity of the predicted value of DTA.
\end{itemize}

\section{Related work}

In recent years, with the remarkable success of artificial intelligence techniques in computer modeling, speech recognition and natural language processing, many regression-based deep learning computational models have been proposed for DTA prediction.
The following methods have shown remarkable results in the related research on the application of DTA prediction using SMILES character sequences of drug molecules and protein target sequences as input.

\subsection{KronRLS}

According to existing proofs, regularized least squares models \cite{8pahikkala2015toward} can predict binary drug-target interactions with high accuracy.
KronRLS can be treated as a generalization of this type of models for predicting continuous binding affinity values.
Given a set ${\{ {y_i}\} _i}_{ = 1}^m$ consisting of drug and target and an association vector containing the affinity relations between them, KronRLS learns a prediction function by minimizing the loss of Eq. 1.

\begin{equation}
\label{eq1}
 {{\cal J}(f) = \sum\limits_{i = 1}^m {({y_i} - f(} {x_i}){)^2} + \lambda ||f||_k^2}
\end{equation}

where $\lambda  > 0$ is the user-defined regularization parameter, $||f||_k^2$ is the norm of $f$, and $k$ is the kernel function.

\subsection{DeepDTA}

DeepDTA \cite{10ozturk2018deepdta} as a deep learning-based model uses sequence information of drugs and targets to predict binding affinity.
First, this method converts drug and target sequences into matrices as network input and uses two convolutional neural network (CNN) blocks to extract drug and target features.
Then, the feature information is input into a feature prediction block composed of several fully connected layers. Finally, the affinity prediction value is output.
The loss function of the model is the mean squared error between the predicted affinity and the true affinity.

\begin{equation}
\label{eq2}
 {{\cal J}_{DeepDTA}} = \frac{1}{n}\sum\limits_{i = 1}^n {{{({{\hat Y}_i} - {Y_i})}^2}}
\end{equation}

where $n$ is the number of samples, ${\hat Y_i}$ is the predicted value, and ${Y_i}$ is the corresponding true affinity value.

\subsection{DeepAffinity}

DeepAffinity \cite{18karimi2019deepaffinity} uses sequence information of drugs and targets to predict drug-target affinity.
This method transforms the recurrent neural network (RNN) model and selects the gated recurrent unit (GRU) \cite{sutskever2014sequence} as the default Seq2Seq model \cite{20li2021co}.
The composite SMILES character sequence is first embedded in a fingerprint, and then the encoder is used to map the input SMILES character sequence to a fixed-dimensional vector. Finally, the decoder maps this vector to the target output.
Deep features of drugs or proteins are computed by a pre-trained unsupervised Seq2Seq autoencoder model, and then appended to the encoder's output by a 1D convolutional layer and a
CNN block composed of a max pooling layer to predict affinity values.

\subsection{GraphDTA}

GraphDTA \cite{12nguyen2021graphdta} uses sequence information and graphical information to represent targets and drugs respectively.
The GraphDTA model first uses the Rdkit \cite{22landrum2013rdkit} toolkit to convert the SMILES character sequences of a drug into a subgraph, and then uses a graph convolutional network (GCN) to extract drug features.
Then, the CNN is used to extract the features of target sequences. Finally, several fully connected layers are used to obtain the predicted affinity.

\subsection{Co-VAE}

Co-VAE \cite{20li2021co} utilizes the sequence information of drug and target to predict drug-target affinity.
This method transforms the traditional VAE \cite{23kingma2013auto} model, which makes the model interpretable to a certain extent while completing the prediction task.
Specifically, the model maps the input SMILES character sequence and target molecule sequence to a fixed-dimensional vector through the encoder respectively, and the decoder maps the two vectors to the target SMILES sequence and target molecule sequence respectively.
The decoder consists of a three-layer gated CNN and several fully connected layers. It processes the output vectors of the two decoders through two progressive fully connected layers. A fully connected layer is added to process the feature information to predict the affinity value.

\subsection{SeqGO-CPA}

SeqGO-CPA \cite{24wang2021seqgo} uses the GO vocabulary system \cite{25carbon2009amigo} to extract the structural feature information of drug molecules and protein molecular sequences to achieve the prediction of drug-target affinity.
Specifically, the model takes SMILES character sequences representing drug molecules, amino acid sequences representing protein molecules, and GO terms of the corresponding proteins as input.
GO annotations are represented by a pre-trained natural language processing model Bert \cite{26devlin2018bert}. The structures of SMILES character sequences and amino acid sequences of protein molecules are decomposed by an NLP tokenization algorithm \cite{27kudo2018sentencepiece}, respectively.
Subsequently, feature information is learned from the above three substructures using CNN blocks and global max pooling.
The final features of drugs and proteins are linked together and fed into three fully linked layers for binding affinity prediction.

\subsection{TransDTI}

TransDTI \cite{13kalakoti2022transdti} developed an end-to-end extensible framework to extract drug-target pair sequence interaction information, and finally achieved the prediction of drug-target affinity.
Specifically, TransDTI uses a Transformer-based language model to extract ordered information presented in datasets such as SMILES and protein sequences to infer interactions between the given drug and the corresponding target.
The model first uses a variety of language models to extract protein molecular sequences and drug SMILES character sequence feature information, and then embeds them in the same dimension vector.
The prediction of interaction scores and interaction states between two sequences is done by training a fully connected feedforward deep neural network on the extracted embedding vectors.
Finally, the performance of the prediction model is verified by using molecular docking and kinetics analysis.

\subsection{MolTrans}

MolTrans \cite{14huang2021moltrans} formulated DTI prediction as a classification task to determine whether a pair of drugs and target proteins interacts.
Given the input drug molecule SMILES character sequence and protein molecule amino acid sequence information, the self-built frequent consecutive sub-sequence (FCS) \cite{14huang2021moltrans} mining module is used to find repeated subsequences in drug and protein databases and decompose them into a group of well-defined substructure sequences.
Enhanced contextual embeddings are then obtained for each substructure through the Transformer \cite{17vaswani2017attention} embedding module.
Pairing of drug substructures with protein substructures with pairwise interaction scores is done by using an interaction prediction module, and high-order interaction information is captured using CNN.
Finally, the decoder module outputs a score indicating the probability of pairwise interaction. DTI prediction is achieved.

\subsection{Summary}

Although the above-mentioned methods can achieve good performance, there are three main deficiencies.

\begin{itemize}
\item
The loss of feature information is caused by different input representations.
Using a one-dimensional sequence to represent the molecular structure as input and directly using tools to extract molecular features cause the loss of feature information.

\item
The interaction mechanism modeling is insufficient.
The binding strength of DTA can be quantified by detecting non-covalent interactions between two molecules.
During this quantification, non-covalent interactions between two molecules result from the sequential binding of atoms and substructures between them. Therefore, the connections between atoms and substructures can contain more characteristic information that affects the binding affinity of drug molecules and protein molecules.
However, the above-mentioned DTA prediction methods cannot characterize the interactions between molecular substructures and between constituent atoms, respectively.

\item
The interpretability is weak. Existing machine learning methods lack a pathway that can indicate the effectiveness of learning a molecular structure or sequence representation of the input.
\end{itemize}

\section{The Proposed method}
The proposed method focuses on completing DTI tasks by creating a mechanism of interactive learning between molecular sequences, so that the prediction network has a certain ability to acquire multi-scale features.
In order to avoid that the molecular interaction information of the prediction network only comes from the simple feature information fusion of two mutually independent molecular sequences, the mechanism of intermolecular interaction and the continuity of affinity measurements are ignored.

\begin{figure}[ht]
	\centering
	\includegraphics[height=6.5cm]{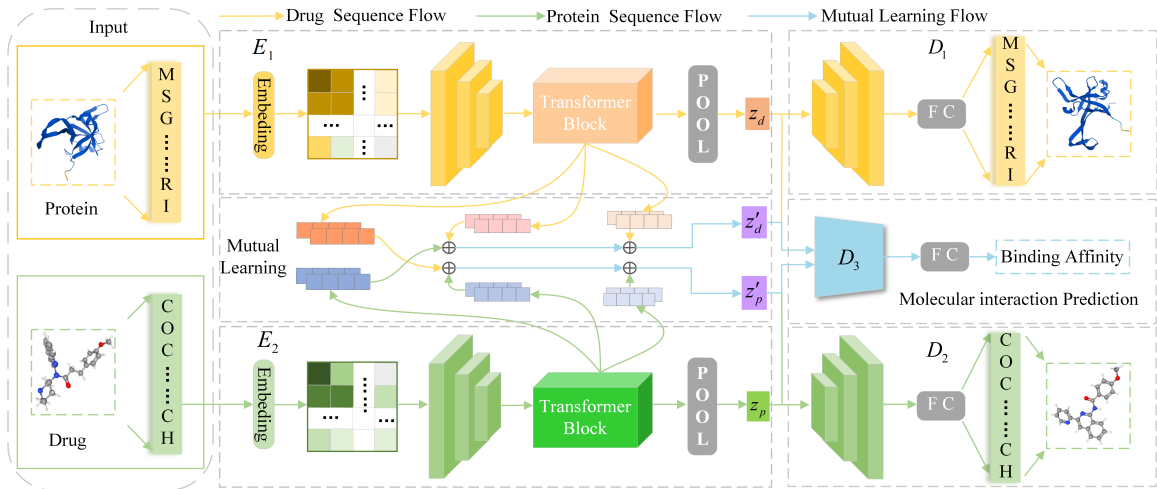}
	\caption{MT-DTA consists of three parts: drug/protein molecule sequence representation learning, information interaction learning between sequences, and drug-target interaction force prediction. Molecular representation learning uses encoders ${E_1}$ and ${E_2}$ to extract multi-scale informative features of molecular sequences, and obtain the string representations of new drug and protein molecules through decoders ${D_1}$ and ${D_2}$ respectively. Information interaction mechanism learning uses Transformer self-attention mechanism to extract interactive information features between molecular sequences. Drug target affinity prediction is used to predict intermolecular affinity values by fusing feature information with decoder $D_3$.}
	\label{fig2}
\end{figure}

Problem definition: Given a set of drug molecular sequences ${X_d} = \{ {x_d}^i\} _{i = 1}^n$ represented by the SMILES character sequence code, where ${x_d}^i$ is the molecular sequence of the i-th in ${X_d}$ , and $n$ is the total number of molecules in ${X_d}$;
a set of protein molecular sequences ${X_p} = \{ {x_t}^i\} _{i = 1}^m$ represented by FASTA \cite{29pearson1988improved} construction, where ${x_p}^i$ is the $i$-th molecular sequence in ${X_p}$, and $m$ is the total number of molecules in ${X_p}$;
a set of binding affinity matrices ${X_b} = \{ {B_i}\} _{i = 1}^n$, where ${B_i}$ represents the $1 \times m$ dimensional affinity matrix with all molecular sequences in ${X_p}$, and a single element in ${B_i}$ is represented as ${K_{i,j}}$ that is the $j$-th affinity value in ${B_i}$.
In order to enhance the feature extraction ability of the network for one-dimensional sequence data and the interpretability of the network, this paper adopts the automatic variational encoder (VAE) model \cite{23kingma2013auto} as the basic network framework.
As shown in Fig. 2, the technical framework of the proposed solution is mainly composed of three parts: drug/protein molecule sequence representation learning, inter-sequence information interaction learning, and drug-target interaction force prediction.
The drug/protein molecular sequence representation learning endows the model with multi-scale feature acquisition capabilities, enabling the model to extract a more complete hidden feature variable distribution of the input sequence \cite{zhao2021transformer}.
Additionally, the loss of molecular feature information caused by local feature extraction is alleviated.
Inter-sequence information interaction learning is responsible for establishing the relation between the feature information of two independent molecular sequences, so as to avoid the degradation of affinity prediction performance caused by ignoring the intermolecular interaction mechanism in the process of model feature extraction.
Drug-target interaction force prediction is used to predict the interaction force between molecular sequences by fusing the feature information of two molecular sequences.

\subsection{Drug/Protein Molecular Sequence Representation Learning}

In the whole process, feature encoders ${E_1}$ and ${E_2}$ are first trained by using a set of drug molecule sequences represented by SMILES codes and a set of protein target molecule sequences.
As shown in Fig. 3, the two encoders ${E_1}$ and ${E_2}$ have the same structure, consisting of a label encoding layer, an embedding layer, a CNN block, and a Transformer block.
The specific structure of representation learning is shown in Fig. 3.
Specifically, the sequence of the drug/protein molecule is first mapped to the atomic number in the dictionary according to the atom/amino acid symbol input code.
Subsequently, the embedding layer converts the input sequence into vectors of the same size.
The downsampling operation converts the sequence into a two-dimensional map, and the degree of overlap can be controlled by the stride size.
Then, it is input into successive convolutional layers to obtain a feature map $\{ {F_{ij}}|i \in n,j \in m\} $ in matrix form, where $F_{ij}$ represents the feature map of the $i$-th drug sequence in $X_d$ and the $j$-th protein sequence in $X_p$.
In order to obtain the global information of molecular sequences, an attention-based transformer \cite{17vaswani2017attention} model is adopted.
In the transformer model, a multi-head attention mechanism, a fully connected feedforward network and a two-layer linear change are included, and the activation function is ReLU \cite{28nair2010rectified}.
The multi-head attention mechanism projects the sequence information onto the three linear spaces of Query, Key and Value, and combines them into different attention results through different linear transformations.
The representation within each attention module is obtained.
Finally, different attention modules are spliced to represent the final output. The corresponding formula is expressed as follows.

\begin{equation}
\label{eq3}
 {{\begin{array}{c}
Attention(Q,K,V) = Softmax(\frac{{Q{K^T}}}{{\sqrt {{d_k}} }})V,\\
hea{d_i} = Attention(QW_i^Q,K{W_i}^K,V{W_i}^V),\\
MultiHead(Q,K,V) = Concat(hea{d_1}, \ldots hea{d_h}).
\end{array}}}
\end{equation}

The matrices $QW_i^Q \in {R^{{d_{model}} \times {{\rm{d}}_k}}}$, $K{W_i}^K \in {R^{{d_{model}} \times {{\rm{d}}_k}}}$ and $V{W_i}^V \in {R^{{d_{model}} \times {{\rm{d}}_k}}}$ represent the corresponding linear projections respectively.
And $\{ d_{model},{d_k} \}$ represent the dimensions of the model and K, respectively.
$h$ is the number of multi-head attention modules and $i$ is the number of heads in each attention module.
Inspired by the encoder structure in the Transformer model, this paper incorporates an attention mechanism into drug/protein molecular sequence representation learning.
Specifically, each element in the output matrix $F_{ij}$ of the convolutional layer is represented as a token \cite{17vaswani2017attention} in the sequence of drug/protein molecules, and it is flattened into the sets ${S_d} = \{ {s_d}^i\} _{i = 1}^{{n_d}}$ and ${S_p} = \{ {s_p}^i\} _{i = 1}^{{n_p}}$ of multiple token sequences.
$\{ {s_d}^i,{s_p}^i\} $ represent the subsequence dictionary mapping value respectively, and $\{ {n_d},{n_p}\} $ represent the number of subsequences respectively.
A regularization token is used for tokenization. Each stage of tokenization is allowed to gradually reduce the number of tokens while increasing the corresponding feature dimension.
The tokenization helps to achieve multi-scale aggregation of different feature information extracted by convolution mechanism and attention mechanism.
The Transformer layer uses a self-attention mechanism to associate the information from the token sequence sets converted from different molecular sequence inputs.
The input sequence is projected into the three subspaces of Query, Key and Value with weights. All token sequence relations are considered to help generate global attention weights.
Additionally, classification tokens are only added in the last stage.
Finally, the multi-layer perceptron (MLP) is used to output the upper-level classification label and the corresponding token sets, and obtain the latent feature variable representation of each category.

\begin{figure}[ht]
	\centering
    \includegraphics[height=6.5cm]{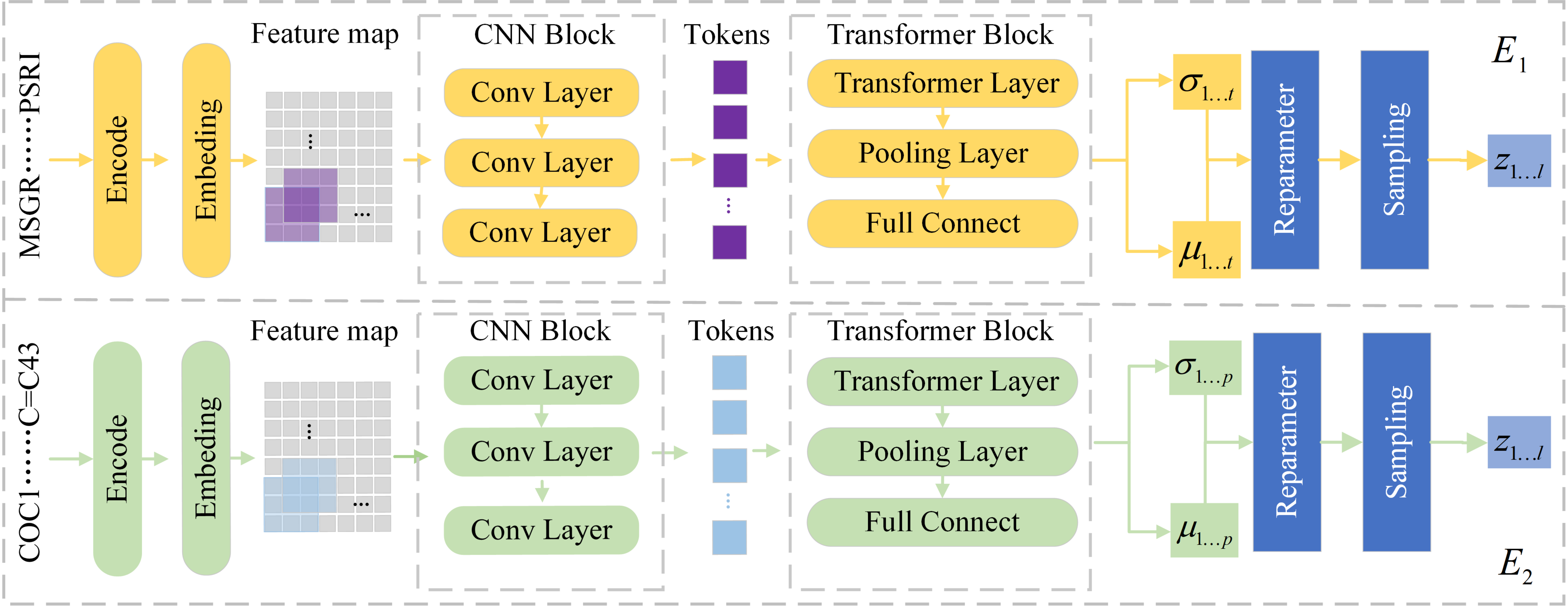}
	\caption{Drug/Protein Molecular Presentation Learning. $E_1$ and $E_2$ represent feature encoders for SMILES character sequences of drug molecules and amino acid sequences of protein molecules respectively. ${\sigma _{1, \ldots l}}$ and ${\mu _{1, \ldots l}}$ represent the $l$-th variance and mean of the output, respectively. ${z_{1, \ldots l}}$ represents the corresponding $l$-th hidden feature variable.}
	\label{fig3}
\end{figure}

\subsection{Information interactive learning mechanism}

Information interaction learning is used to capture fine-grained feature information between molecules. Transformer's self-attention mechanism builds global connections of molecular sequences by learning all subsequences from the same molecular sequence. Since the patch token in the transform layer has the ability of information interaction, the self-attention mechanism of the two transformer blocks is established as a mutual learning mechanism in this paper, as shown in Fig. 4.

\begin{figure}[ht]
	\centering
	\includegraphics[height=10cm]{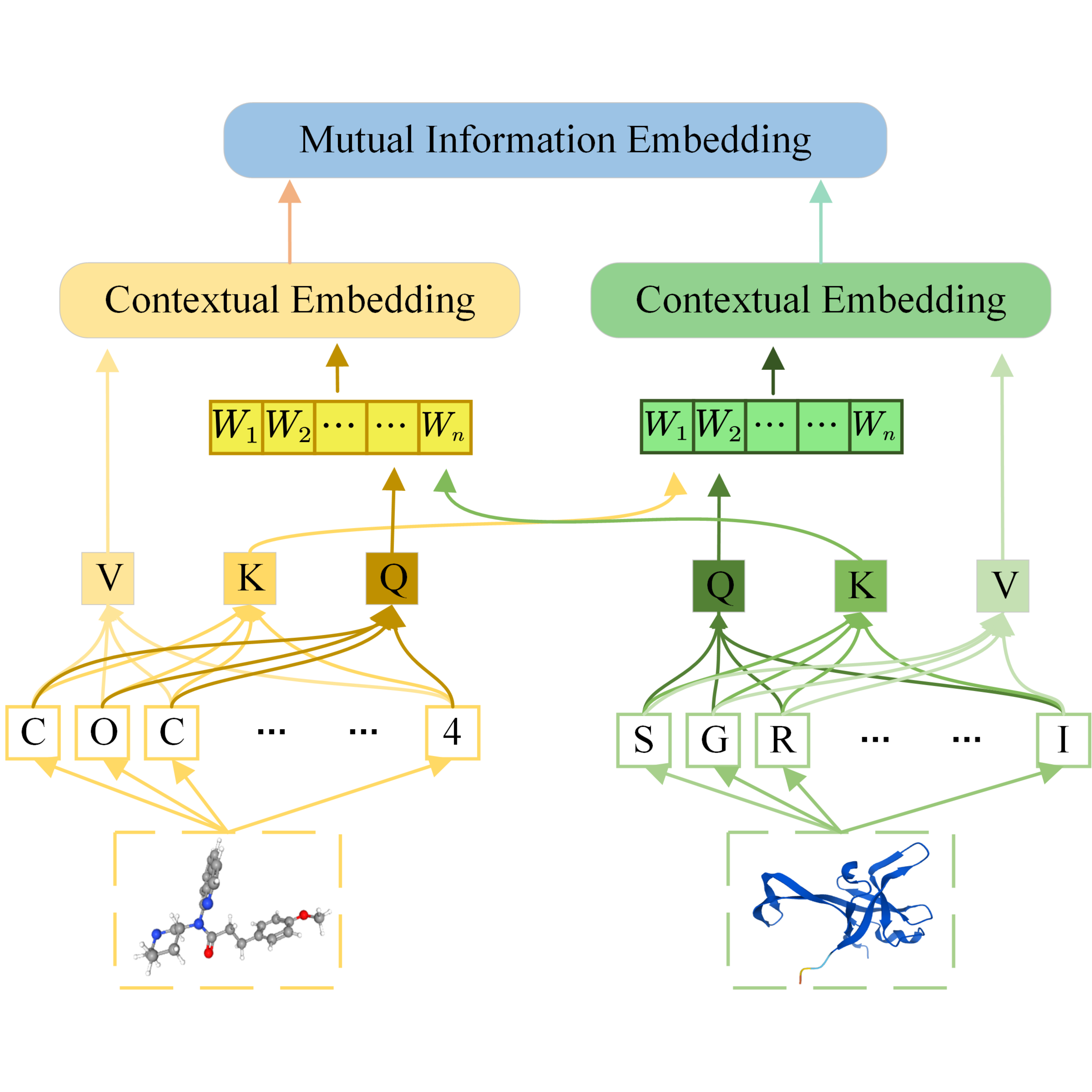}
	\caption{Information interaction learning. The left and right parts respectively represent the subsequence self-attention learning of the SMILES character sequence of drug molecules and the amino acid sequence of protein molecules. Q, K and V are the corresponding Query, Key and Value linear spaces respectively, $\{ {W_i}\} _{i = 1}^N$ represents the network self-learning weight.}
	\label{fig4}
\end{figure}

Specifically, the attention functions of the two molecular sequences shown in Eq. 4 are used to transform the multi-head attention.

\begin{equation}
\label{eq4}
 {f({Q_d},{K_d},{V_d}) = \left\{ {\begin{array}{*{20}{c}}
{{W_{{q_d}}}{S_d}}\\
{{W_{{k_d}}}{S_d}}\\
{{W_{{v_d}}}{S_d}}
\end{array},} \right.f({Q_p},{K_p},{V_p}) = \left\{ {\begin{array}{*{20}{c}}
{{W_{{q_p}}}{S_p}}\\
{{W_{{k_p}}}{S_p}}\\
{{W_{{v_p}}}{S_p}}
\end{array},} \right.}
\end{equation}

where $\{ f({Q_d},{K_d},{V_d}),f({Q_p},{K_p},{V_p})\} $ is denoted as the attention function of drug molecule sequence and protein molecule sequence, respectively, the input $\{ {S_d},{S_p}\} $ is linearly projected into subspaces $Q_d$, $K_d$, $V_d$ and $Q_p$, $K_p$, $V_p$ respectively, and the corresponding weights are ${W_{{q_d}}}$, ${W_{{k_d}}}$, ${W_{{v_d}}}$ and ${W_{{q_p}}}$, ${W_{{k_p}}}$, ${W_{{v_p}}}$ respectively.

As shown in Eq. 5, they are connected by using a multi-head attention mechanism.

\begin{equation}
\label{eq5}
 {{C_m} = {\rm{ }}Softmax\{ (\frac{{{Q_d}K_p^T}}{{\sqrt {{d_d}} }}){V_d},(\frac{{{Q_p}K_d^T}}{{\sqrt {{d_p}} }}){V_p}\} }
\end{equation}

where $C_m$ is the $1 \times n$ dimensional attention feature matrix, and $\{ {d_D},{d_P}\} $ respectively represent the dimension of the molecular sequence $\{ {S_d},{S_p}\} $.
Eq. 6 takes into account the interaction information of intermolecular subsequences, thereby helping to generate the global attention vector between molecules.
In the calculation process of self-attention, the aggregation of the token sequence is shown as follows.

\begin{equation}
\label{eq6}
 {{F_{avg}} = \frac{{\sum\limits_{i = 1}^{n + 1} {{C_m}(i,j)} }}{{n + 1}}}
\end{equation}

where ${F_{avg}}$ denotes the aggregation features of drug molecules and protein molecules, $i$ and $j$ denote the rows and columns of ${C_m}$ respectively.

Finally, the aggregated features are reprojected back into the self-attention mechanism as shown in Eq. 7.

\begin{equation}
\label{eq7}
 {O = {Concat}(({\eta _1}[{W_{F{C_1}}}{\eta _2}({W_{F{C_2}}}{F_{avg}})])}
\end{equation}

where $\{ {W_{F{C_1}}},{W_{F{C_2}}}\} $ respectively represent the weights of the two fully connected layers, $\{ {\eta _1},{\eta _2}\}$ respectively represent the ReLU and the linear activation function, $O$ represents the aggregation result of molecular sequence features containing interaction information. The probability score of each Token is obtained, which is mapped to multiple output elements.

Similarly \cite{20li2021co}, the short-cut structure in the residual network is adopted to solve the degradation problem in deep learning.
$O$ is converted into the mean parameter $\mu $ and the variance parameter $\sigma $ through the residual network.
Then, the output hidden feature variable z is obtained by sampling with the reparameterization trick techniques.

\subsection{Drug-target affinity prediction}

The drug target binding affinity prediction training decoder $D_3$ first aggregates the input features of the two encoders to obtain the aggregated features, and then re-projects them into the multi-head attention mechanism. After the normalization layer, the final output is a scalar corresponding to the elements in the affinity matrix.
Specifically, the decoder $D_3$ first receives the output features of the interactive learning module, splits them and pools the maximum value through a pooling operation.
Discriminative information on the role of drug molecules and protein molecular substructures is obtained.
The pooled max value is then fed to the fully connected layer, and the two outputs are concatenated into a vector, which is fed to the fully connected layer to aggregate into a tensor.
The drug molecule feature tensor is attached to their respective protein molecule feature tensor. The token that forms the drug-target pair includes the interactions between the drug molecule and the various substructures of the protein molecule.
The location encoding is then added to this Token to help Transformer obtain and associate relative location information.
This feature information is input into the multi-head attention module, and the output label vector reflects the interaction information of the drug-target pair.
Intramolecular neighboring substructures and intermolecular substructures interact when triggering interactions. The normalization layer generates a single scalar that unambiguously measures the strength of interactions between individual pairs of target drug substructures.
In particular, for the encoders $E_1$ and $E_2$, this paper sets the decoders $D_1$ and $D_2$, which are similar in structure, and uses deconvolution to reconstruct the encoder input features.
Each deconvolution operation corresponds to the convolution operation of the encoder, and finally the prediction sequence is output through the fully connected layer.
Specifically, the structure of $E_1$ and $E_2$ contains two fully connected layers and three deconvolution layers.
Finally, the new drug sequence encoded by SMILES character sequence or the new protein molecule sequence encoded by FASTA is output.

The implementation of the MT-DTA model has two main steps. Each step uses the pseudocode segment with the largest execution volume as the measure of the time complexity of the algorithm in this step.
In step one, the pseudocode segment with the largest execution range is a for loop, and its time complexity is denoted as $O(n)$.
The largest pseudocode segment executed in the second step is a while loop. Specifically, the loop judgment condition is whether the model converges, which is a constant level, and the time complexity is denoted as $O(1)$.
In the loop body, the maximum complexity of the calculation step is linear, and its time complexity is denoted as $O(m)$.
Then the time complexity of step two is $O(m)$.
Therefore, the total complexity of the model algorithm is $O(n+m)$, and the proposed MT-DTA algorithm is summarized as follows.

\begin{breakablealgorithm}
\setstretch{1.5}
	\caption{MT-DTA Prediction Model Processing}
	\small
	\begin{algorithmic}[1]
		\Require
		$x_d$ is the drug SMLES string as model input; $x_p$ is the protein target sequence string as input; $s_d$ and $s_p$ are the set of token sequences corresponding to drug SMLES string and protein molecular sequence respectively; $y$ is the true value of the binding affinity of $x_d$ and $x_p$.
		\Ensure
		 $\hat{x}_d$ and $\hat{x}_p$ are the predicted values corresponding to the input $x_d$ and $x_p$ respectively; $\hat{y}$ is the predicted values corresponding to the input $y$.
        \State {\bf{Step1.}}
		\For {$i = 1...n$}{$// n $ is the number of input.\strut}
            \State $ (s_d^i,s_p^i), lexical$\,$order \leftarrow select(s_d,s_p) $
            \If {lexical order $\in$ numbers format}
                \State call Assignment$(s_d^i,s_p^i)$ $//$convert atomic name into numbers
            \Else
                 \State break
            \EndIf
        \EndFor
        \State {\bf{Step2.}}
        \State $i \leftarrow 0$
		\While {not convergence of objective}
		\State Sample $\{x_d^i\}_{i=1}^n $ and $\{x_p^i\}_{i=1}^n$ from data distribute $p_{{\theta}_d}(x_d)$ and $p_{{\theta}_p}(x_p)$.
        \State Sample $\{z_d^i\}_{i=1}^n $ and $\{z_p^i\}_{i=1}^n$ from data distribute $p_{{\theta}_d}(z_d)$ and $p_{{\theta}_p}(z_p)$.
        \State Sample $\{{\epsilon}_d^i\}_{i=1}^n $ and $\{{\epsilon}_p^i\}_{i=1}^n$ from $\mathcal{N}(0,1)$.
		\State Compute $\theta$-gradient (Eq.26)
		\State $g_{\theta}\leftarrow \frac{1}{n}\sum_{i=1}^{n}\nabla_{\theta} \log p_{\theta}(x_d^i|z_{\theta}(x_d^i|\epsilon_{d}^{i}))$
		\State Compute $\phi$-gradient (Eq.26)
		\State $g_{\phi}\leftarrow \frac{1}{n}\sum_{i=1}^{n}\nabla_{\phi} \log p_{\phi}(x_p^i|z_{\phi}(x_p^i|\epsilon_{p}^{i}))$
		\State Compute $\hat{y}$-gradient (Eq.26)
		\State $\hat{y}\leftarrow \frac{1}{n}\sum_{i=1}^{n}\nabla_{\theta,\phi} \log p_{\theta,\phi}(x_d^i,x_p^i|z_{\theta,\phi}\{(x_d^i,{\epsilon}_d^i),(x_p^i,{\epsilon}_p^i)\})$
		\State Perform Adam \cite{23kingma2013auto} updates for $\theta$ and $\phi$ :
		\State $ \theta \leftarrow \theta$ + $h_ig_{\theta} $,$ \phi \leftarrow \phi$ + $h_ig_{\phi} $,$ \hat{y} \leftarrow \hat{y}$ + $h_i(g_{\theta},g_{\phi}) $,
		\State $ i\leftarrow i$ + $1$
		\EndWhile
	\State Calculate average loss to determine various hyperparameters
	\State \Return the trained model

	\end{algorithmic}
\label{alg1}
\end{breakablealgorithm}

\section{ELBO Maximization Proof for MT-DTA Model}

VAE \cite{23kingma2013auto} as a generative model is developed using deep learning methods based on statistical theory.
Given a set $X = \{ {x^i}\} _{i = 1}^N$ of multidimensional variables consisting of random variables $x^i$, VAE can generate a variable $\hat{x}$ similar to the original $x$ by learning the distribution of the hidden space $Z$ of $X$.
VAE can be divided into two parts: a recognition model ${q_\phi }(z|x)$ and a generative model ${p_\theta }(z|x)$.
For the purpose of VAE, the marginal likelihood lower bound $\log {p_\theta }({x^i})$ is maximized by using a neural network to learn the parameters $\phi $ and $\theta $ of the two models.
Specifically, the input variable $x$ obtains the hidden feature variable $z$ of $x$ through the recognition model ${q_\phi }(z|x)$.
The distribution of sample $z$ is used as input, and the predictor variable $\hat{x}$ is obtained through the generative model ${p_\theta }(x|z)$.
The VAE loss function objective can be expressed as Eq. 8.

\begin{equation}
\label{eq8}
 {log\;{p_\theta }({x^i}) \ge {\cal L}(\theta ,\phi ;{x^i}) = {\mathbb{E}_{{q_\phi }(z|{x^i})}}[log\;{p_\theta }({x^i}|z)] - {D_{KL}}({q_\phi }(z|{x^i})||{p_\theta }(z)]}
\end{equation}
where ${D_{KL}}({q_\phi }(z|{x^i})||{p_\theta }(z))$ is the Kullback-Leibler(KL) divergence with respect to ${q_\phi }(z|{x^i})$ and ${p_\theta }(z)$, and ${\cal L}(\theta ,\phi ;{x^i})$ is ELBO.
The symbol "$||$" connects two different distributions and $\mathbb{E}$ represents mathematical expectations.

As a drug target binding affinity prediction problem, this paper assumes random variables of drug molecule sequence $x_d$, and random variables of protein molecular sequence $x_p$, and random variables of affinity value $y$.
$x_d$ and $x_p$ are represented by SMILES character sequences and protein molecule sequences respectively, and $y$ is a positive real number.
There are two input sets $D = \{ {x_d}^1,{x_d}^2, \ldots ,{x_d}^i\} _{i = 1}^N$ and $P = \{ {x_p}^1,{x_p}^2, \ldots ,{x_p}^i\} _{i = 1}^N$.
As the model goal, a predictive model $p(y|{x_d},{x_p})$ is learned by training a triplet $\{ x_d^i,x_p^i,{y^i}\} _{i = 1}^N$.
The hidden feature variables of $x_d$ and $x_p$ are assumed to be $z_d$ and $z_p$ respectively, and they are independent of each other.
${{S}_{d}}=\{{{s}_{d}}^{i}\}_{i=1}^{{{n}_{d}}}$ and ${{S}_{p}}=\{{{s}_{p}}^{i}\}_{i=1}^{{{n}_{p}}}$are the sets of subsequences of the corresponding sequence respectively, where $\{{{s}_{d}}^{i},{{s}_{p}}^{i}\}$are the corresponding subsequence dictionary mapping values (represented by integers), and $\{{{n}_{d}},{{n}_{p}}\}$ represents the number of subsequences.
${{z}_{s}}$ represents the hidden feature vector of the set of subsequences.
The corresponding probabilistic relation between the MT-DTA model structures is presented by the drug/protein molecular representation learning correspondence recognition models ${{q}_{{{\phi }_{d}}}}({{z}_{d}}|{{x}_{d}})$ and ${{q}_{{{\phi }_{p}}}}({{z}_{p}}|{{x}_{p}})$, which generate hidden feature variables ${{z}_{d}}$ and ${{z}_{p}}$ corresponding to the input ${{x}_{d}}$ and ${{x}_{p}}$.
The decoders ${{D}_{1}}$, ${{D}_{2}}$, and the drug-target affinity prediction module are prediction models ${{p}_{{{\theta }_{d}}}}({{x}_{d}}|{{z}_{d}})$, ${{p}_{{{\theta }_{p}}}}({{x}_{p}}|{{z}_{p}})$ and ${{p}_{{{\theta }_{b}}}}(y|{{z}_{d}},{{z}_{p}})$ respectively, which produce distributions $p({{x}_{d}})$, $p({{x}_{p}})$ and $p(y)$ corresponding to the inputs ${{x}_{d}}$, ${{x}_{p}}$ and $\{{{x}_{d}},{{x}_{p}}\}$ \cite{23kingma2013auto}.
Based on the assumptions, it can be proved that the model optimization focuses on maximizing the lower bound of the evidence for the joint distribution of the triplet $\{x_{d}^{i},x_{p}^{i},{{y}^{i}}\}$.
Since the hidden feature variables of ${{x}_{d}}$ and ${{x}_{p}}$ are ${{z}_{d}}$ and ${{z}_{p}}$ respectively, their priors
are ${{p}_{{{\theta }_{d}}}}({{z}_{d}})$ and ${{p}_{{{\theta }_{p}}}}({{z}_{p}})$ respectively.
The lower bounds of the log-likelihood of the distribution prediction model can be formalized as follows.

\begin{equation}
\label{eq9}
\begin{array}{l}
log\;{p_{{\theta _d}}}(x_d^i) \ge {\cal L}(\theta ,\phi ;x_d^i)\\
 = {\mathbb{E}_{{q_{{\phi _d}}}({z_d}|x_d^i)}}[log\;{p_{{\theta _d}}}(x_d^i|{z_d})] - {D_{KL}}({q_{{\phi _d}}}({z_d}|x_d^i)||{p_{{\theta _d}}}({z_d}));\\
log\;{p_{{\theta _p}}}(x_p^i) \ge {\cal L}(\theta ,\phi ;x_p^i)\\
 = {\mathbb{E}_{{q_{{\phi _p}}}({z_p}|x_p^i)}}[log\;{p_{{\theta _p}}}(x_p^i|{z_p})] - {D_{KL}}({q_{{\phi _p}}}({z_p}|x_p^i)||{p_{{\theta _p}}}({z_p}));\\
log\;{p_{{\theta _b}}}({y^i}) \ge {\cal L}(\theta ,\phi ;{z_d},{z_p})\\
 = {\mathbb{E}_{{q_\phi }({z_d},{z_p}|{x_d}^i,{x_p}^i,{y^i})}}\log {p_{{\theta _b}}}({x_d}^i,{x_p}^i,{y^i}|{z_d},{z_p})\\
 - {D_{KL}}({q_{{\phi }}}({z_d},{z_p}|{x_d}^i,{x_p}^i,{y^i})||{p_{{\theta _b}}}({z_d},{z_p}))
\end{array}
\end{equation}

The lower bound on the log-marginal likelihood of the joint distribution of the combined triplet $\{x_{d}^{i},x_{p}^{i},{{y}^{i}}\}$ is shown as follows.

\begin{equation}
\label{eq10}
\begin{array}{l}
log\;{p_\theta }(x_d^i,x_p^j,{y^i}) \ge {\cal L}(\theta ,\phi ;x_d^i,x_p^i,{y^i}) = \\
\{ {\mathbb{E}_{{q_{{\phi _d}}}({z_d}|x_d^i)}}[log\;{p_{{\theta _d}}}(x_d^i|{z_d})] - {D_{KL}}({q_{{\phi _d}}}({z_d}|x_d^i)||{p_{{\theta _d}}}({z_d})\} \\
 + \{ {\mathbb{E}_{{q_{{\phi _p}}}({z_p}|x_p^i)}}[log\;{p_{{\theta _p}}}(x_p^i|{z_p})] - {D_{KL}}({q_{{\phi _p}}}({z_p}|x_p^i)||{p_{{\theta _P}}}({z_p})\} \\
 + \{ {\mathbb{E}_{{q_\phi }({z_d},{z_p}|x_d^i,x_p^i)}}[log\;{p_{{\theta _b}}}({y^i}|{z_d},{z_p})] - {D_{KL}}({q_\phi }({z_d},{z_p}|x_d^i,x_p^i)||{p_{{\theta _b}}}({y^i}|{z_d},{z_p}))\}
\end{array}
\end{equation}

where $\theta =\{{{\theta }_{d}},{{\theta }_{p}},{{\theta }_{b}}\}$, $\phi =\{{{\phi }_{d}},{{\phi }_{p}}\}$.

Proof: According to Eq. 8, the lower bound of the log-likelihood of the joint distribution$\{x_{d}^{i},x_{p}^{i},{{y}^{i}}\}$ can be formalized as follows.

\begin{equation}
\label{eq11}
\begin{array}{l}
\log {p_\theta }({x_d}^i,{x_p}^i,{y^i}) \ge {\cal L}(\theta ,\phi ;{x_d}^i,{x_p}^i,{y^i})\\
 = {\mathbb{E}_{{q_\phi }({z_d},{z_p}|{x_d}^i,{x_p}^i,{y^i})}}\log {p_\theta }({x_d}^i,{x_p}^i,{y^i}|{z_d},{z_p}) \\
 - {D_{KL}}({q_{{\phi}}}({z_d},{z_p}|{x_d}^i,{x_p}^i,{y^i})||{p_\theta }({z_d},{z_p}))
\end{array}
\end{equation}

where the variational distribution ${{q}_{\phi }}({{z}_{d}},{{z}_{p}}|{{x}_{d}}^{i},{{x}_{p}}^{i},{{y}^{i}})$ is an approximation of the posterior distribution ${{p}_{\theta }}({{z}_{d}},{{z}_{p}}|{{x}_{d}}^{i},{{x}_{p}}^{i},{{y}^{i}})$.

In the interaction of drug and protein molecular information, each input sequence is divided into multiple token sequences $\{{{s}_{d}}^{i},{{s}_{p}}^{i}\}_{i=1}^{N}$.
The corresponding high-dimensional hidden feature vector $\{{{z}_{ds}},{{z}_{ps}}\}\in {{R}^{*}}$ is generated by MLP, and the final output $\{{{z}_{d}},{{z}_{p}}\}$ is obtained. The corresponding process is shown as follows.

\begin{equation}
\label{eq12}
{q_\phi }({z_{ds}},{z_{ps}}|s_d^i,s_p^i) = {q_{{\phi _d}}}({z_{ds}}|s_d^i){q_{{\phi _p}}}({z_{ds}}|s_p^i)
\end{equation}

Since the hidden feature vector $\{{{z}_{ds}},{{z}_{ps}}\}$ constitutes the hidden feature vector $\{{{z}_{d}},{{z}_{p}}\}$  $\{{{z}_{ds}},{{z}_{ps}}\}$ and $\{{{z}_{d}},{{z}_{p}}\}$, and the prior distribution ${{p}_{{{\theta }_{d}}}}({{z}_{ds}})$ and ${{p}_{{{\theta }_{p}}}}({{z}_{ps}})$ are independent of each other, Eq. 13 is obtained.

\begin{equation}
\label{eq13}
\sum\limits_{i = 1}^N {{q_\phi }({z_{ds}},{z_{ps}}|s_d^i,s_p^i)}  \approx {q_\phi }({z_d},{z_p}|x_d^i,x_p^i)
\end{equation}

Furthermore, the variational distribution ${q_{\phi}}$ and the posterior distribution ${p_{\theta}}$ in Eq. 11 can be simplified according to the multivariate Bayesian formula as follows.

\begin{equation}
\label{eq14}
\begin{array}{c}
{q_\phi }({z_d},{z_p}|{x_d^i},{x_p^i},{y^i}) = {q_{{\phi _d}}}({z_d}|{x_d^i}){q_{{\phi _p}}}({z_p}|{x_p^i}), \\
{p_\theta }({z_d},{z_p}|{x_d^i},{x_p^i},{y^i}) = {p_{{\theta _d}}}({x_d^i}|{z_d}){p_{{\theta _p}}}({x_p^i}|{z_p}){p_{{\theta _b}}}({y^i}|{z_d},{z_p}).
\end{array}
\end{equation}

Then the first term of Eq. 11 can be simplified as Eq. 15.

\begin{equation}
\label{eq15}
\begin{array}{l}
{\mathbb{E}_{{q_\phi }({z_d},{z_p}|{x_d^i},{x_p^i},{y^i})}}\log {p_\theta }({x_d^i},{x_p^i},{y^i}|{z_d},{z_p}) \\
= {\mathbb{E}_{{q_{{\phi _d}}}({z_d}|{x_d^i})}}\log {p_{{\theta _d}}}({x_d^i}|{z_d}) + {\mathbb{E}_{{q_{{\phi _p}}}({z_p}|{x_p^i})}}\log {p_\theta }({x_p^i}|{z_p}) \\
 + {\mathbb{E}_{{q_\phi }({z_d},{z_p}|{x_d^i},{x_p^i})}}\log {p_{{\theta _b}}}({y^i}|{z_d},{z_p})
\end{array}
\end{equation}

Due to the mutual information, the prior distributions ${{p}_{{{\theta }_{d}}}}({{z}_{d}})$ and ${{p}_{{{\theta }_{p}}}}({{z}_{p}})$ are not independent of each other, as shown in Eq. 16.

\begin{equation}
\label{eq16}
{p_\theta }({z_d},{z_p}) = {p_\theta }[{p_{{\theta _d}}}({z_d})|{p_{{\theta _p}}}({z_p})]{p_{{\theta _p}}}({z_p})
\end{equation}

Then the second term of Eq. 11 can be simplified as Eq. 17.

\begin{equation}
\label{eq17}
\begin{array}{l}
{D_{KL}}({q_{{\phi}}}({z_d},{z_p}|{x_d^i},{x_p^i},{y^i})||{p_\theta }({z_d},{z_p})) \\
= {D_{KL}}({q_{{\phi _d}}}({z_d}|{x_d^i})||{p_{{\theta _d}}}({z_d})) + {D_{KL}}({q_{{\phi _p}}}({z_p}|{x_p^i})||{p_{{\theta _p}}}({z_p}))\\
 - {D_{KL}}({q_\phi }({z_d},{z_p}|x_d^i,x_p^i)||{p_{{\theta _b}}}({y^i}|{z_d},{z_p}))
\end{array}
\end{equation}

According to Eqs. 15 and 17, Eq. 11 is then simplified as Eq. 18.

\begin{equation}
\label{eq18}
\begin{array}{l}
\log {p_\theta }({x_d^i},{x_p^i},{y^i}) \ge {\cal L}(\theta ,\phi ;{x_d^i},{x_p^i},{y^i})\\
 = {\mathbb{E}_{{q_{{\phi _d}}}({z_d}|{x_d^i})}}\log {p_{{\theta _d}}}({x_d^i}|{z_d}) - {D_{KL}}({q_{{\phi _d}}}({z_d}|{x_d^i})||{p_{{\theta _d}}}({z_d}))\\
 + {\mathbb{E}_{{q_{{\phi _p}}}({z_p}|{x_p^i})}}\log {p_\theta }({x_p^i}|{z_p}) - {D_{KL}}({q_{{\phi _p}}}({z_p}|{x_p^i})||{p_{{\theta _p}}}({z_p}))\\
 + {\mathbb{E}_{{q_\phi }({z_d},{z_p}|{x_d^i},{x_p^i})}}\log {p_{{\theta _b}}}({y^i}|{z_d},{z_p}) + {D_{KL}}({q_\phi }({z_d},{z_p}|x_d^i,x_p^i)||{p_{{\theta _b}}}({y^i}|{z_d},{z_p}))
\end{array}
\end{equation}

Let the recognition model distribution ${{p}_{{{\theta }_{d}}}}({{x}_{d}^{i}},{{z}_{d}})$ and ${{p}_{{{\theta }_{p}}}}({{x}_{p}^{i}},{{z}_{p}})$ be the classification distribution, and the prior distribution ${{p}_{{{\theta }_{d}}}}({{z}_{d}})$ and ${{p}_{{{\theta }_{p}}}}({{z}_{p}})$ of the hidden feature variables ${{z}_{d}}$ and ${{z}_{p}}$ obey the Gaussian distribution $\mathcal{N}(0,I)$.
According to the high-dimensional characteristics of the input random variables, let the variational distribution ${{q}_{{{\phi }_{d}}}}({{z}_{d}}|{{x}_{d}^{i}})$ and ${{q}_{{{\phi }_{p}}}}({{z}_{p}}|{{x}_{p}^{i}})$ be a multivariate Gaussian distribution, and the logarithmic form is expressed as Eq. 19.

\begin{equation}
\label{eq19}
\begin{array}{l}
\log {q_{{\phi _{{\rm{ *}}}}}}({z_{\rm{*}}}|x_*^i) = \log {\cal N}(\mu _*^i,{(\sigma _*^i)^2}I)
\end{array}
\end{equation}

where $*=\{d,p\}$, mean $\mu _{*}^{i}$ and variance $\sigma _{*}^{i}$ are the outputs of the recognition model corresponding to $ x_{*}^{i}$.
The dependent variational distribution ${{q}_{{{\phi }_{d}}}}({{z}_{d}}|{{x}_{d}^{i}})$ is an approximation of the posterior distribution ${{p}_{{{\theta }_{d}}}}({{x}_{d}^{i}}|{{z}_{d}})$, which is shown as follows.

\begin{equation}
\label{eq20}
\begin{array}{l}
\log {p_{{\theta _*}}}(x_*^i|{z_*}) = \log {\cal N}(x_*^i,\mu _*^i,{(\sigma _*^i)^2}I)
\end{array}
\end{equation}

According to the re-parameter techniques \cite{20li2021co}, Eq. 21 is obtained.

\begin{equation}
\label{eq21}
\begin{array}{l}
z_{d}^{i,m} = \mu_{i}^{d} + \sigma_{i}^{d} \odot \varepsilon _{d}^{m}, z_{p}^{i,m} = \mu_{i}^{p} + \sigma_{i}^{p} \odot \varepsilon _{p}^{m}.
\end{array}
\end{equation}

where $z_{d}^{i,m}$ and $z_{p}^{i,m}$represent the $m$-th samples of the hidden feature variables $z_{d}^{i}$ and $z_{p}^{i}$ respectively, $\varepsilon _{d}^{m}$ and $\varepsilon _{p}^{m}$ are the $m$-th re-parameters respectively, $\varepsilon _{d}^{m} \sim \mathcal{N}(0,I)$ and $\varepsilon _{p}^{m} \sim \mathcal{N}(0,I)$.
And $\odot$ is represented the multiplication operation.
Since the neural network under the multivariate Gaussian distribution has a diagonal covariance structure, the mean $\mu$ and variance $\sigma$ under the structure of this paper can be solved by the following formula.

\begin{equation}
\label{eq22}
\left\{ \begin{array}{c}
\mu  = {W_1}h + {b_1},\\
\log {\sigma ^2} = {W_2}h + {b_2},\\
h = MultiHead(c,{L_{out}}(x),{L_{out}}(x)).
\end{array} \right.
\end{equation}

where $\text{ }\!\!\{\!\!\text{ }{{W}_{1}},{{W}_{2}}\}$ an d$\text{ }\!\!\{\!\!\text{ }{{b}_{1}},{{b}_{2}}\}$ are the weight and bias parameters of the model network respectively, and $h$ is the multi-head attention output obtained by merging the context vector $c$, and ${{L}_{out}}(x)$ indicates that the encoder takes the input as the output of $x$.
According to Eqs. 20 ~ 22, the posterior distributions ${{({{x}_{d}^{i}})}^{T}}\log {{\hat{x}}_{d}^{i}}$ and ${{({{x}_{p}^{i}})}^{T}}\log {{\hat{x}}_{p}^{i}}$ of generative models $\log {{p}_{{{\theta }_{d}}}}({{x}_{d}^{i}}|{{z}_{d}})$ and $\log {{p}_{{{\theta }_{p}}}}({{x}_{p}^{i}}|{{z}_{p}})$ are obtained respectively. The posterior distribution of the generative model $\log {{p}_{{{\theta }_{b}}}}({{y}^{i}}|{{z}_{d}},{{z}_{t}})$ is shown as follows.

\begin{equation}
\label{eq23}
\log {p_{{\theta _b}}}({y^i}|{z_d},{z_t}) =  - \beta {({y^i} - {\hat y^i})^2} + C.
\end{equation}

where $\{{{\hat{x}}_{d}^{i}},{{\hat{x}}_{p}^{i}},{{\hat{y}}^{i}}\}$ is the reconstruction of ${{\hat{x}}_{d}^{i}}$, ${{\hat{x}}_{p}^{i}}$ and ${{y}^{i}}$ respectively, $\beta$ is the regularization parameter, and $C$ is a constant.
The MT-DTA model is used to maximize the lower bound of the logarithm-likelihood of the joint distribution $\{x_{d}^{i},x_{p}^{i},{{y}^{i}}\}$. According to Eqs. 11 $\sim$ 23, Eq. 24 is obtained.

\begin{equation}
\label{eq24}
\begin{array}{l}
{\cal L}(\theta ,\phi ;{x_d^i},{x_p^i},{y^i})\\
 \approx \frac{1}{{{L_d}}}\sum\limits_{{l_d} = 1}^{{L_d}} {[({{({x_d^i})}^T}\log {{\hat x}_d^i})} (z_d^{i,l}) + \frac{1}{2}\sum\limits_{{j_d} = 1}^{{J_d}} {(1 + \log {{(\sigma _{d,{j_d}}^i)}^2} - {{(\mu _{d,{j_d}}^i)}^2} - {{(\sigma _{d,{j_d}}^i)}^2})} \\
 + \frac{1}{{{L_p}}}\sum\limits_{{l_p} = 1}^{{L_p}} {[({{({x_p^i})}^T}\log {{\hat x}_p}^i)} (z_p^{i,l}) + \frac{1}{2}\sum\limits_{{j_p} = 1}^{{J_p}} {(1 + \log {{(\sigma _{p,{j_p}}^i)}^2} - {{(\mu _{p,{j_p}}^i)}^2} - {{(\sigma _{p,{j_p}}^i)}^2})} \\
 - [\beta \sum\limits_{{l_d} = 1}^{{L_d}} {\sum\limits_{{l_p} = 1}^{{L_p}} {({{({y^i} - {{\hat y}^i})}^2} + C)} } \\
 - \frac{1}{2}\sum\limits_{{j_d} = 1}^{{J_d}} {\sum\limits_{{j_p} = 1}^{{J_p}} {[(1 + \log {{(\sigma _{d,{j_d}}^i\sigma _{p,{j_p}}^i)}^2} - {{(\mu _{d,{j_d}}^i)}^2} - {{(\mu _{p,{j_p}}^i)}^2} - {{(\sigma _{d,{j_d}}^i)}^2} - {{(\sigma _{p,{j_p}}^i)}^2}} } ]
\end{array}
\end{equation}

where $\{{{J}_{d}},{{J}_{p}}\}$ are the dimensions of the hidden feature variables $\{{{z}_{d}},{{z}_{p}}\}$ respectively, $\{{{L}_{d}},{{L}_{p}}\}$ are the sampling numbers of the posterior distributions ${{p}_{{{\theta }_{d}}}}({{x}_{d}^{i}}|{{z}_{d}})$ and ${{p}_{{{\theta }_{p}}}}({{x}_{p}^{i}}|{{z}_{p}})$ respectively, $\{\sigma _{d,{{j}_{d}}}^{i},\sigma _{p,{{j}_{p}}}^{i},\mu _{d,{{j}_{d}}}^{i},\mu _{p,{{j}_{p}}}^{i}\}$ respectively represent the ${{j}_{*}}$-th element in the vector $\{\sigma _{d}^{i},\sigma _{p}^{i},\mu _{d}^{i},\mu _{p}^{i}\}$, and $*\in \{d,p\}$.

Therefore, the evidence lower bound of the joint distribution $\{x_{d}^{i},x_{p}^{i},{{y}^{i}}\}$ can be maximized by the MT-DTA model, as shown in Eq. 24.

\section{Experiments and Results}
In order to make the performance test of the prediction model more robust and accurate, the random cross-comparison method is used as the experimental setting \cite{18karimi2019deepaffinity} to predict the affinity of the test group drug/target to all the targets/drugs in the dataset.
The details are shown as follows: (1) the drugs/targets in the dataset are randomly divided into six equal groups, (2) five groups are randomly selected from them as the training group, and the remaining group is used as the test group, (3) drugs/targets are randomly selected ten times under the above settings, and the mean and standard deviation of all evaluation metrics are measured.
To avoid the problem of tensor size asymmetry: in the Davis dataset, the drug SMILES character sequence and the target protein sequence are embedded in matrices with dimensions (85, 128) and (1200, 128) respectively;
in the KIBA dataset, the drug SMILES character sequences and target protein sequences are embedded in matrices with dimensions (100, 128) and (1000, 128) respectively.
The proposed MT-DTA was implemented on Pytorch 1.10.2, and all the experiments and field tests were conducted on a desktop with 24.00GB NVIDIA GeForce GTX 3090, Intel Core i7-8700MQ CPU @ 3.20GHz and 48.00GB memory.

\subsection{Experimental dataset}

This paper evaluates the performance of the MT-DTA model on two large-scale Kinase Inhibitor Biochemical Metrics Measurement Datasets Davis and KIBA respectively.
The Davis dataset contains 68 drugs and 442 targets, and uses the kinase dissociation constant  to represent the binding affinity between the drug and the target.
The value of $K_d$ reflects the binding tightness of the drug and the target. The value of $K_d$ increases, the binding tightness also increases. The value range of $K_d$ is [0.016, 10000].
In order to make the experimental results more intuitive and easier to compare, this paper converts $K_d$ to logarithmic space $pK_d$, which is expressed as follows.

\begin{equation}
\label{eq25}
p{K_d} =  - {\log _{10}}(\frac{{{K_d}}}{{{{10}^9}}})
\end{equation}

The frequency histogram of the affinity relation represented by the value of $pK_d$ is shown in Fig. 5(a).

\begin{figure}[ht]
	\centering
	\includegraphics[height=5cm]{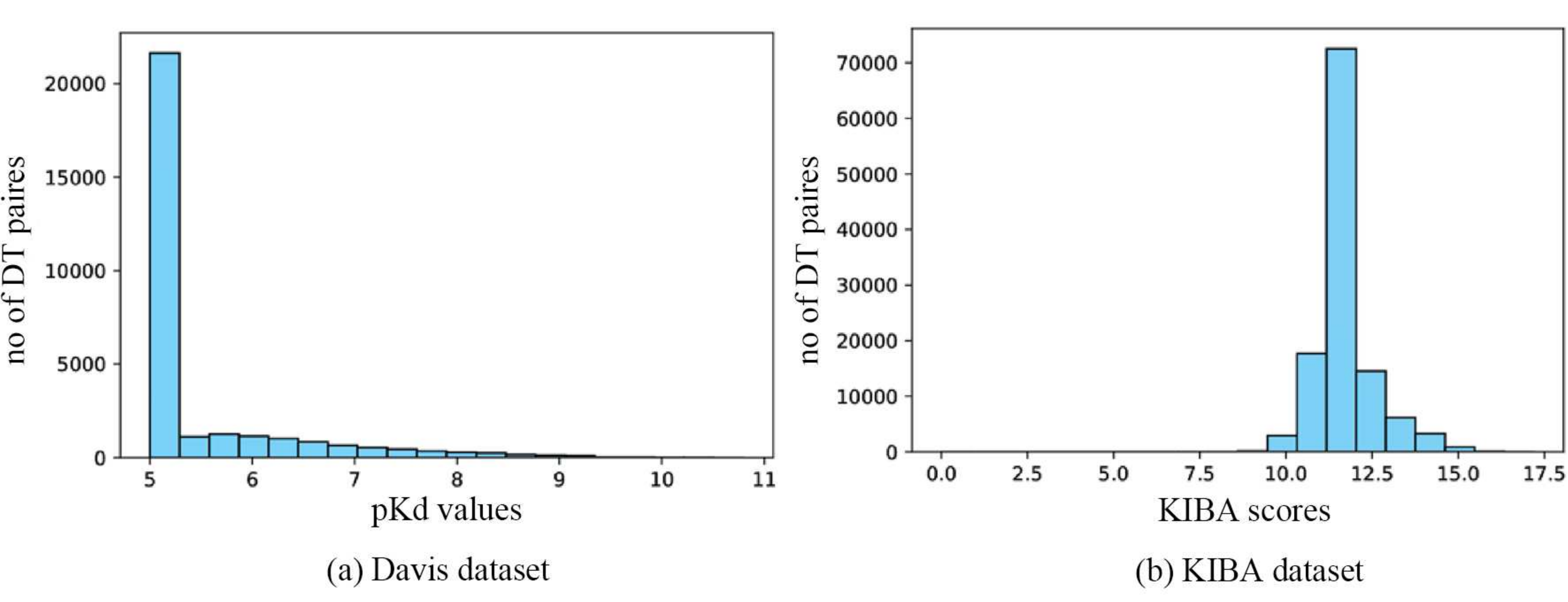}
	\caption{Histogram of affinity relation frequency for (a) Davis dataset and (b) KIBA dataset}
	\label{fig5}
\end{figure}

The KIBA dataset integrates biological activity values under the action of different kinase inhibitors into a biological activity matrix.
Binding affinities in this matrix are given as KIBA values.
The value of KIBA score increases, the binding tightness between the drug and the target decreases, and the value range of KIBA is [0, 17.2].
By removing drug-target pairs with score values less than 10 in the KIBA dataset, a new dataset containing 2,116 drugs and 229 targets is obtained with a density of 24.4$\%$.
The dataset is binarized by using a threshold with a value of KIBA $\leqslant$ 3.0 \cite{20li2021co}, and then the threshold becomes 12.1 in the transformed dataset.
The frequency histogram of affinity relation represented by KIBA scores is shown in Fig. 5(b).
The length distributions of SMILES character sequences and protein sequences for drugs in the Davis and KIBA datasets are shown in Fig. 6(a) and Fig. 6(b), respectively.

\begin{figure}[ht]
	\centering
	\includegraphics[height=10cm]{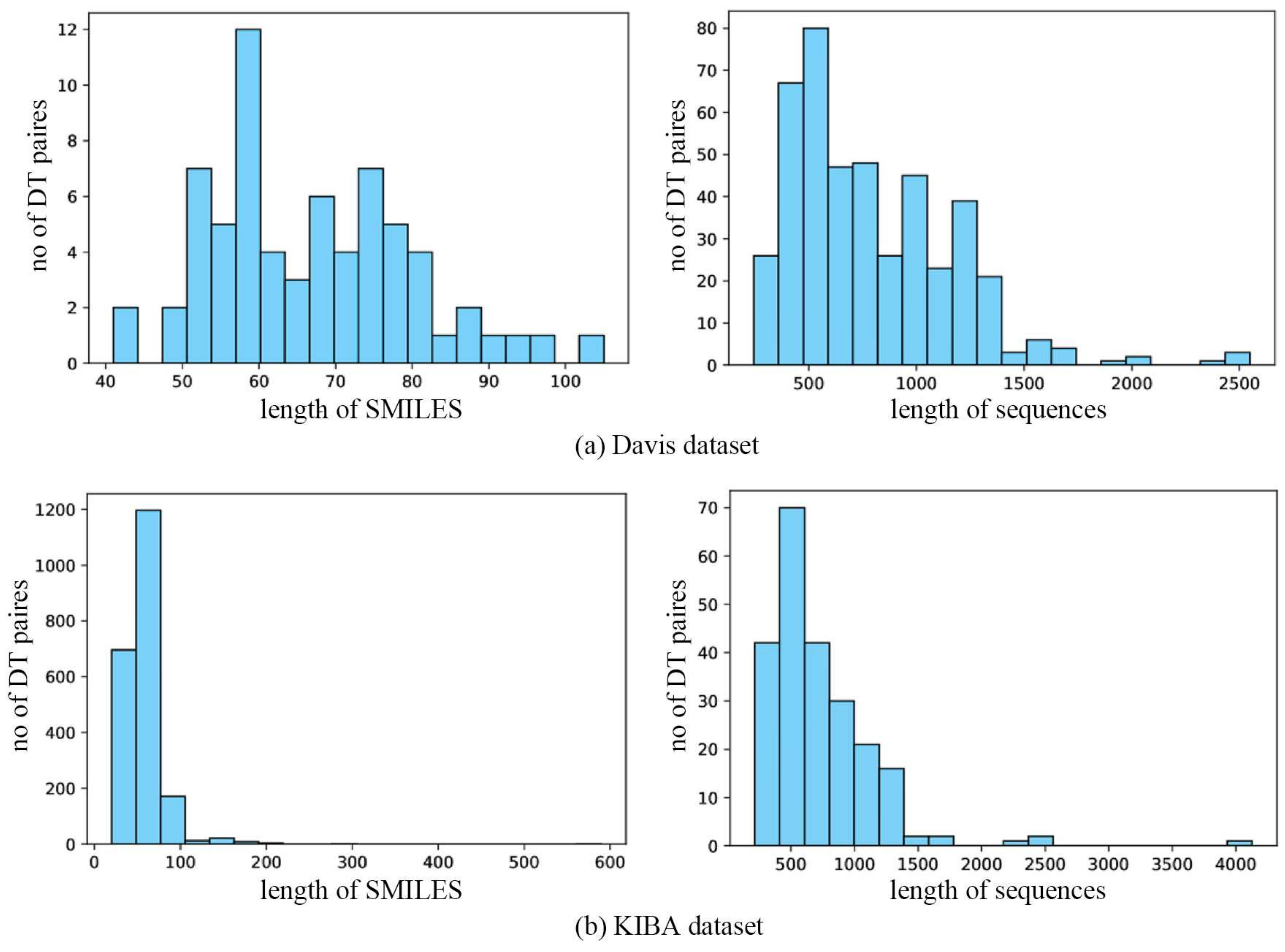}
	\caption{Histograms of length-frequency histograms of SMILES character sequences and protein molecule sequences for (a) Davis dataset and (b) KIBA dataset}
	\label{fig6}
\end{figure}

\subsection{Experimental parameter settings}

The parameter settings of the proposed model \cite{20li2021co} are shown in Tab. 1, including the filter settings in the encoder, the filter settings in the decoder, and the dropout rate.
Specifically, the filter parameter settings in the encoders   and  are 32*1, 32*2, and 32*3 respectively, and the filter parameter settings in the corresponding decoders are 32*3, 32*2, and 32*1 respectively.
The drug SMILES character sequence and target sequence filter size are selected from [5, 7] and [7, 11] respectively. In the two datasets, the maximum length of the SMILES character sequence is 85 and 100 respectively, and the maximum length of the target molecule sequence character sequence is 1200 and 1000 respectively.
Regularization parameter $\lambda $ is chosen from [-3,-5].
During the training process, this paper sets the epoch value to 100, the batch size to 256, and the Adam \cite{23kingma2013auto} algorithm with a learning rate size of 0.001 is used to update the network parameters.

\begin{table}[!htp]\centering
\newcommand{\tabincell}[2]{\begin{tabular}{@{}#1@{}}#2\end{tabular}}
\caption{Experimental parameters}\label{tab: 1}
\scriptsize
\begin{tabular}{ccc}\toprule
\tabincell{c}{Parameters} &\tabincell{c}{range} \\\cmidrule{1-2}
\tabincell{c}{Number of filters in encoder} & \tabincell{c}{32*1;32*2;32*3} \\
\tabincell{c}{Number of filters in decoder} & \tabincell{c}{32*3;32*2;32*1} \\
\tabincell{c}{Drug SMILES strings filters length} & \tabincell{c}{[5,7]} \\
\tabincell{c}{Drug SMILES strings max length(in Davis)} & \tabincell{c}{85} \\
\tabincell{c}{Drug SMILES strings max length(in KIBA)} & \tabincell{c}{100} \\
\tabincell{c}{Target strings filters length} & \tabincell{c}{[7,11]} \\
\tabincell{c}{Target strings max length(in Davis)} & \tabincell{c}{1200} \\
\tabincell{c}{Target strings max length(in KIBA)} & \tabincell{c}{1000} \\
\tabincell{c}{Regularization parameters} & \tabincell{c}{[-3,-5]} \\
\tabincell{c}{The rate of dropout} & \tabincell{c}{0.2} \\
\tabincell{c}{Epoch} & \tabincell{c}{100} \\
\tabincell{c}{Batch size} & \tabincell{c}{256} \\
\tabincell{c}{Learning rate} & \tabincell{c}{0.001} \\
\bottomrule
\end{tabular}
\end{table}

Since Co-VAE, DeepDTA, KronRLS, DeepAffinity, and GraphDTA are all predictive DTA methods based on neural network computation rather than structural design, they are used as the baseline methods in this paper.
They are evaluated using the Davis and KIBA datasets. The performance of MT-DTA and the above methods is compared.
This paper uses mean square error (MSE) \cite{30marmolin1986subjective}, mean absolute error (MAE) \cite{31willmott2005advantages}, consistency index (CI) \cite{32farris1989retention}, modified squared correlation coefficient ($r_m^2$) \cite{33dutilleul1993modifying} and area under the ROC curve (AUC) \cite{34lobo2008auc} five indocators to evaluate the performance of these regression-based models.
The values of MSE and MAE decrease, the model prediction accuracy increases. When the CI value is getting close to 1, the model prediction accuracy becomes better. The value $r_m^2$ increases, the model acceptance rate increases. The AUC value increases, the model performance improves.
In addition, the MT-DTA model can also generate new drugs under the conditions of a given drug.
To evaluate the performance of the model for generating drug sequences, two evaluation indicators, validity and uniqueness \cite{20li2021co}, are used in this paper.
Effectiveness is defined as the proportion of effective drug SMILES character sequences in all generated drug SMILES character sequences. The higher proportion means the better performance.
Uniqueness is defined as the proportion of new drugs in all effective drugs. Similarly, the higher proportion means the better performance.

\subsection{Experiment analysis}

In order to verify the effectiveness of the proposed method, this paper compares the KronRLS, DeepDTA, DeepAffinitiy, GraphDTA and Co-VAE methods on the Davis and KIBA datasets respectively.
The MT-DTA method is first evaluated by plotting a scatterplot of the true and predicted DTA values.
Figs. 7 and 8 show the scatter plots of true and predicted affinities for the Davis and Kiba datasets, respectively.
As the ideal condition, a yellow straight line x would be generated in the scatterplot.
Therefore, the more points concentrated on the line mean the better performance of the corresponding method.
The Davis dataset is shown in Fig. 7. In the drug prediction setting, the points of DeepDTA appear the most scattered around the line, as shown in method (b) in Fig. 7.
The points of MT-DTA appear to be most concentrated around the straight line, as shown in method G in Fig. 7.
DeepAffinity generates the most discrete points, KronRLS generates some outliers below the straight line, and Co-VAE and MT-DTA tend to generate the most concentrated points with few outliers.
For the KIBA dataset, as shown in Fig. 8, KronRLS generates the most discrete points, and most of them are below the straight line.
DeepDTA, DeepAffinity and Co-VAE tend to generate more outliers, and MT-DTA generates the most concentrated points.
Through the scatter plot of the two figures, the performance of the six methods can be visually compared. The MT-DTA method has fewer discrete points, and the distribution of points is more concentrated on a straight line. Therefore, its overall performance is the best.

The performance of MT-DTA method in drug target affinity prediction is evaluated by four quantitative indicators CI, MSE, MAE and $r_m^2$.
Table 2 shows the mean and standard deviation of the four evaluation indicators on the Davis dataset for the six comparative methods.
Under the Target setting, the MT-DTA method achieves the best overall performance among the four indicators, with CI slightly higher than KronRLS.
For the KIBA dataset, the mean and standard deviation under the four evaluation indicators for the six comparative methods are shown in Tab. 2.
Under the Drug setting, the MT-DTA method achieves the best performance on two indicators (CI, rm), and its performance is slightly lower than Co-VAE on the other two indicators (MSE, MAE).
Under the Target setting, the MT-DTA method achieves the best performance on three indicators (CI, MSE, and rm) and the second best performance on the remaining indicator (MAE).
In particular, for the Target setting in the KIBA dataset, the performance of the KronRLS method is significantly lower than that of other deep learning methods.

In addition, the performance of the MT-DTA method in predicting drug-target interactions is further evaluated by the evaluation indicator AUC.
Drug-target pairs are labeled positive and negative respectively by selecting a threshold for true affinity values.
AUC values are then calculated from the labels and predicted affinities.
The KIBA dataset is divided into two different settings by choosing different thresholds in (10.5, 12.5) and the corresponding AUC values are reported in Fig. 9.
According to Fig. 9, the proposed method MT-DTA is more stable than the other five methods on both the Drug setting and the Target setting.
However, when the cutoff value under the Drug setting is greater than 12.00, DeepDTA performs slightly better than MT-DTA.
DeepAffinity performs slightly better than MT-DTA with a small-range cutoff between [11.50, 11.75] under the Target setting.
Overall, Tab. 3 reports the average test results of all comparative methods and MT-DTA under the AUC evaluation metric.

\begin{table}[!htp]\centering
\caption{Comparison of the results obtained by existing methods and the proposed MT-DTA using the objective evaluation indicators CI, MSE, MAE, and $r_m^2$ on Davis and KIBA datasets respectively. The best results are marked in bold.}\label{tab:2}
\scriptsize
\begin{tabular}{lccccccc}\toprule
Dataset &Settings &Method &CI(std) &MSE(std) &MAE(std) & {$r_m^2$} \\\cmidrule{1-7}
\multirow{12}{*}{Davis} &\multirow{6}{*}{Drug} &KronRIS &0.656(0.033) &0.796(0.231) &0.606(0.086) &0.143 \\
& &DeepDTA &0.671(0.023) &0.726(0.091) &0.527(0.033) &0.122 \\
& &DeepAffinity &0.673(0.057) &0.955(0.140) &0.639(0.040) &0.114 \\
& &GraphDTA &0.732(0.067) &0.840(0.058) &0.570(0.020) &\textbf{0.176} \\
& &Co-VAE &0.712(0.063) &0.724(0.096) &0.550(0.048) &0.107 \\
& &MT-DTA(proposed) &\textbf{0.743(0.010)} &\textbf{0.433(0.018)} &\textbf{0.422(0.041)} &0.158 \\\cmidrule{2-7}
&\multirow{6}{*}{Target} &KronRIS &0.836 (0.014) &0.429 (0.055) &0.428 (0.025) &0.448 \\
& &DeepDTA &0.780 (0.071) &0.490 (0.095) &0.456 (0.047) &0.433 \\
& &DeepAffinity &0.803 (0.006) &0.477 (0.019) &0.434 (0.050) &0.444 \\
& &GraphDTA &0.778 (0.016) &0.457 (0.009) &0.397 (0.011) &0.429 \\
& &Co-VAE &0.816 (0.011) &0.416 (0.004) &0.401 (0.007) &0.358 \\
& &MT-DTA(proposed) &\textbf{0.844(0.003)} &\textbf{0.384(0.005)} &\textbf{0.368(0.020)} &\textbf{0.450} \\\midrule
\multirow{12}{*}{KIBA} &\multirow{6}{*}{Drug} &KronRIS &0.729 (0.005) &0.449 (0.005) &0.457 (0.003) &0.355 \\
& &DeepDTA &0.728 (0.007) &0.466 (0.022) &0.419 (0.011) &0.342 \\
& &DeepAffinity &0.707 (0.016) &0.526 (0.024) &0.491 (0.023) &0.283 \\
& &GraphDTA &0.728 (0.008) &0.442 (0.012) &0.447 (0.019) &0.379 \\
& &Co-VAE &0.742 (0.011) &0.416 (0.004) & 0.401 (0.007) &0.358 \\
& &MT-DTA(proposed) &\textbf{0.752(0.008)} &\textbf{0.408(0.007)} & \textbf{0.401(0.009)} &\textbf{0.381} \\\cmidrule{2-7}
&\multirow{6}{*}{Target} &KronRIS &0.591 (0.025) &0.825 (0.042) &0.603 (0.010) &0.399 \\
& &DeepDTA &0.722 (0.017) &0.430 (0.016) &0.421 (0.013) &0.114 \\
& &DeepAffinity &0.700 (0.007) &0.490 (0.016) &0.470 (0.011) &0.324 \\
& &GraphDTA &0.658 (0.050) &0.519 (0.045) &0.470 (0.011) &0.314 \\
& &Co-VAE &0.741 (0.010) &0.421 (0.010) &0.396 (0.005) &0.356 \\
& &MT-DTA(proposed) &\textbf{0.751(0.009)} &\textbf{0.390(0.013)} &\textbf{0.393(0.009)} &\textbf{0.406} \\
\bottomrule
\end{tabular}
\end{table}

\begin{figure}[ht]
	\centering
	\includegraphics[width=15cm]{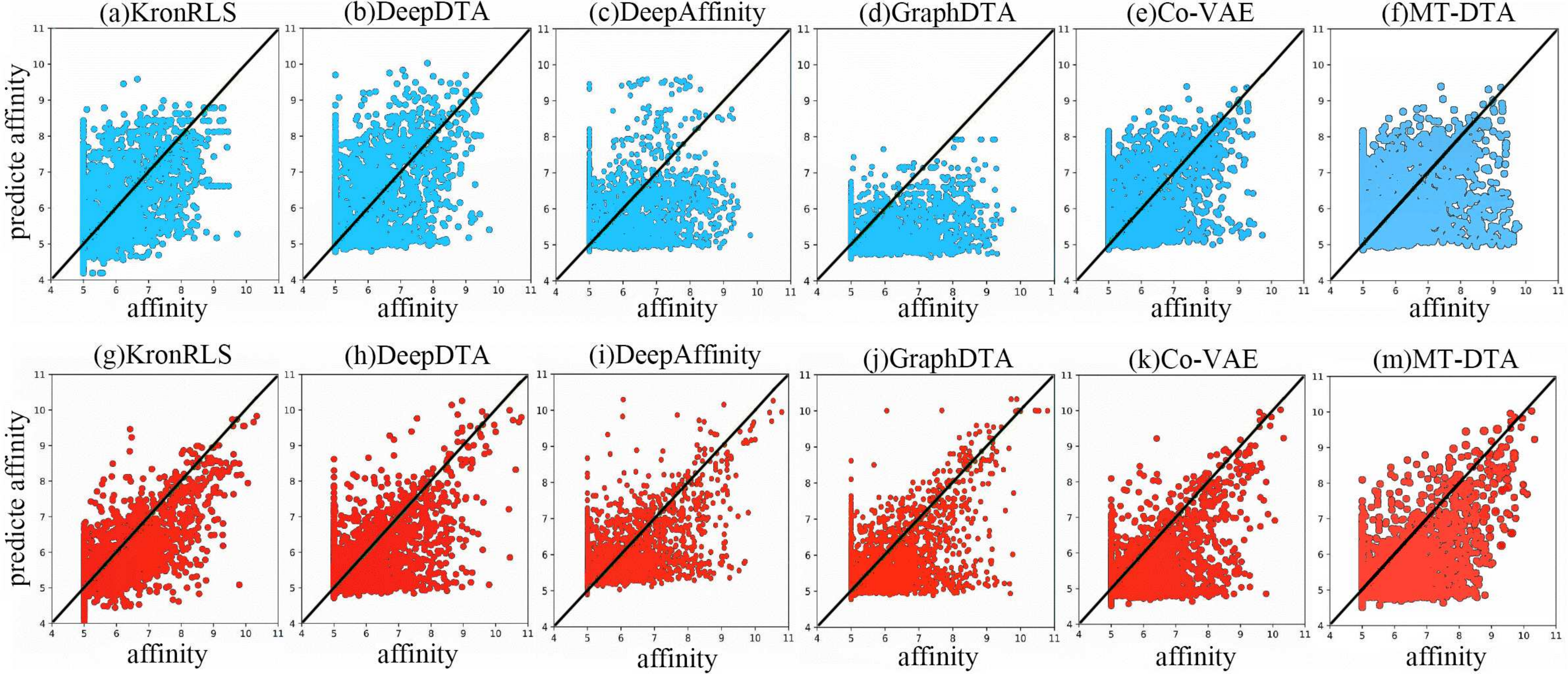}
	\caption{Scatter plots and regression lines of the actual and predicted values of affinity for each method under the Drug(a~f) and Target(g~m) experimental settings respectively. The y-axis is the predicted value of affinity, the x-axis is the true value of affinity, and the dataset is Davis.t}
	\label{fig7}
\end{figure}

\begin{figure}[ht]
	\centering
	\includegraphics[width=15cm]{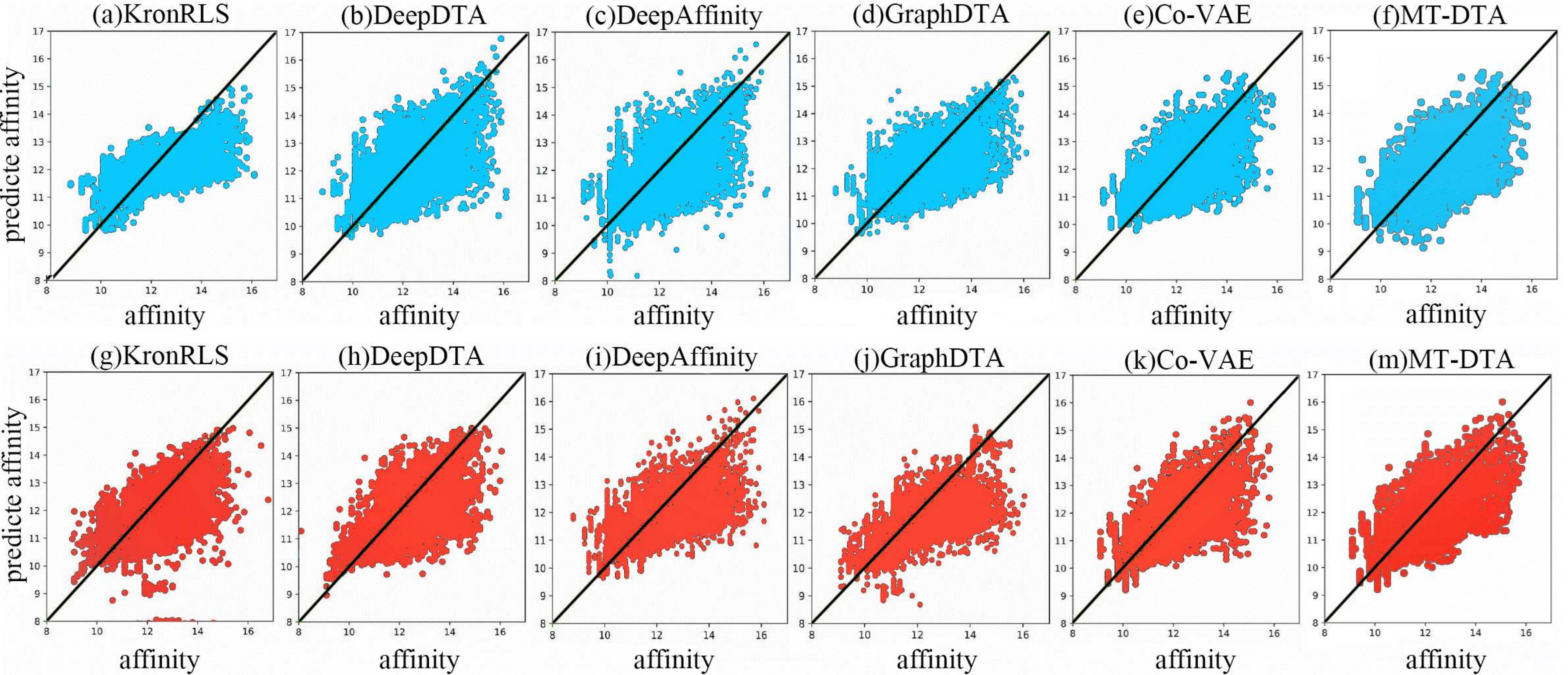}
	\caption{ Scatter plots and regression lines of the actual and predicted values of affinity for each method under the Drug(a~f) and Target(g~m) experimental settings respectively. The y-axis is the predicted value of affinity, the x-axis is the true value of affinity, and the dataset is KIBA.}
	\label{fig8}
\end{figure}

\begin{figure}[htbp]
\label{fig:9}
\subfigure[AUC-Drug]{
\includegraphics[width=7.5cm]{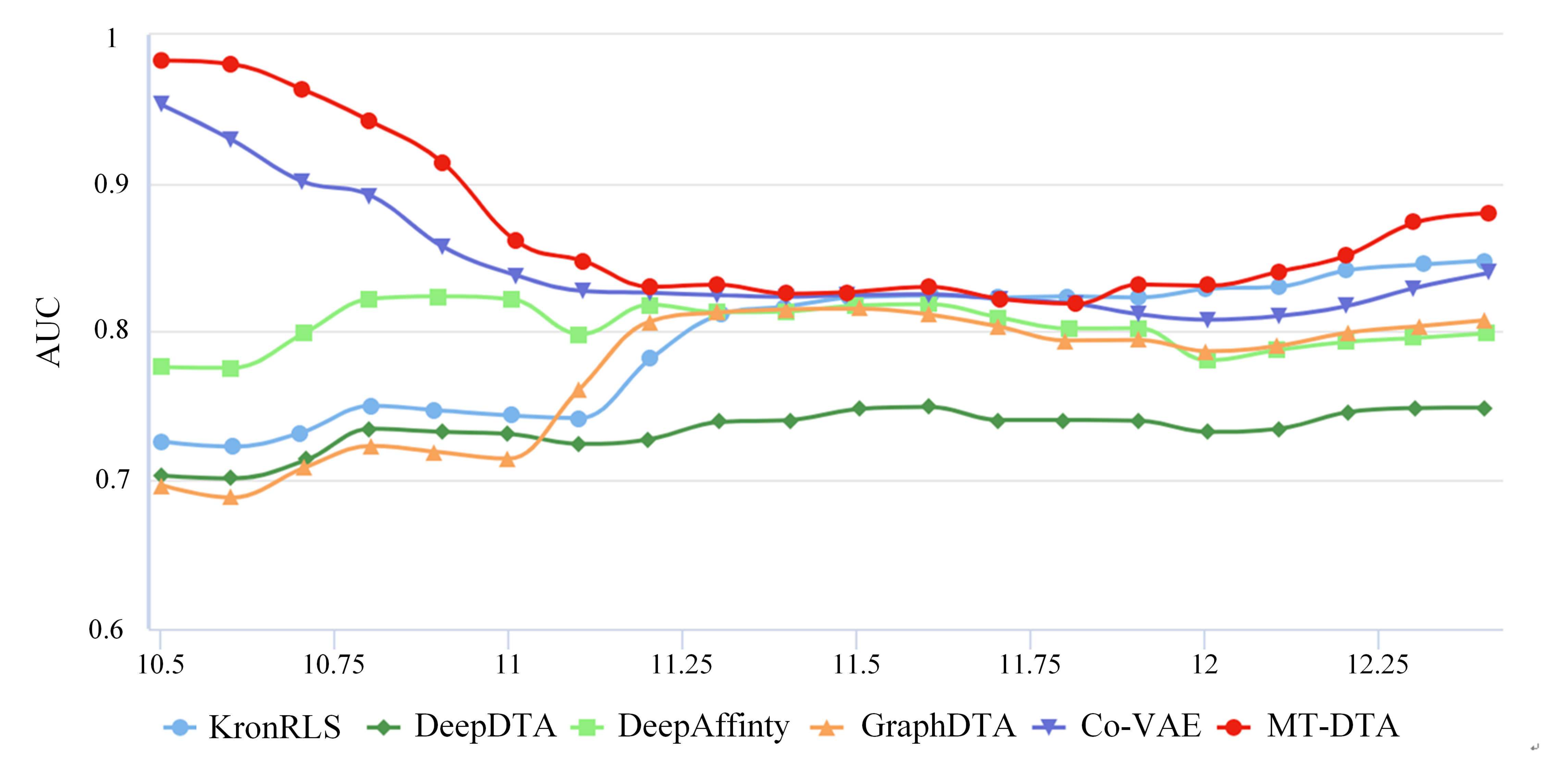}
}
\subfigure[AUC-Target]{
\includegraphics[width=7.5cm]{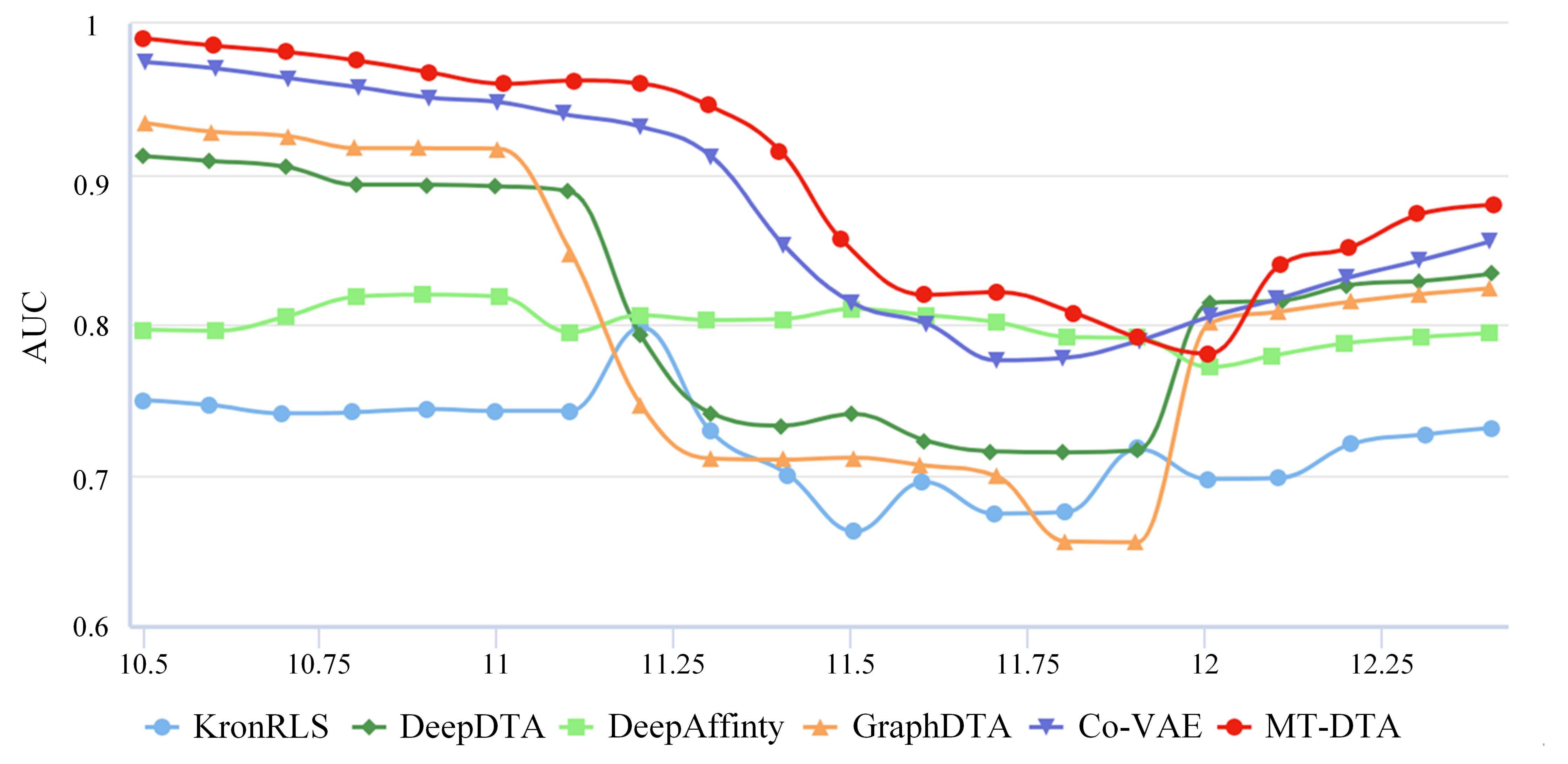}
}
\caption{The before-and-after comparison of the VAD shows that the crying unit in the above figure is monitored and some of the mute fragments are removed. }
\end{figure}

When selecting drug molecule sequences from the KIBA dataset as the input set, the new SMILES character sequences generated by decoder reduction are expected to have similar affinities to the model's input drugs.
This paper evaluates the generative performance of the MT-DTA model by two evaluation indicators validity and uniqueness.
SMILES evaluates effectiveness by calculating percentage of drugs by Rdki \cite{22landrum2013rdkit} and uniqueness by calculating the percentage of new effective drugs not in the KIBA dataset.
The MT-DTA model is compared with Co-VAE, VAE, AAE \cite{35kadurin2017drugan}, GAN \cite{36goodfellow2014generative}, and seqGAN \cite{20li2021co}, and the results are shown in Tab. 3.
The MT-DTA model achieves the best performance on the evaluation indicators of effectiveness and uniqueness.
Finally, some examples of generated SMILES character sequences for drugs are given in Tab. 4, and the examples of generated SMILES character sequences can be found in the KIBA dataset.
Their CIDs and corresponding targets can be found in the database PubChem.
According to Tab. 4, the generated drug SMILES character sequence is usually extremely similar to the original input drug SMILES character sequence with only a small change.
For example, when the SMILES character sequence of the drug CID 49830947 is used as input, the MT-DTA model generates a new SMILES character sequence with CID 11993789 that only changes the last element of the input SMILES sequence.

\begin{table}[!htp]\centering
\newcommand{\tabincell}[2]{\begin{tabular}{@{}#1@{}}#2\end{tabular}}
\caption{Comparison of the results of GAN, SeqGAN, AAE, VAE, Co-VAE and MT-DTA on the effectiveness and uniqueness of evaluation indicators on the KIBA dataset}\label{tab: }
\scriptsize
\begin{tabular}{cccc}\toprule
\tabincell{c}{Model} &\tabincell{c}{Validity(\%)} & \tabincell{c}{Uniquencess(\%)} \\\cmidrule{1-3}
\tabincell{c}{GAN} &\tabincell{c}{<1} &\tabincell{c}{-} \\
\tabincell{c}{SeqGAN} &\tabincell{c}{<1} &\tabincell{c}{-} \\
\tabincell{c}{AAE} &\tabincell{c}{1.26} &\tabincell{c}{0.69} \\
\tabincell{c}{VAE} &\tabincell{c}{48.54} &\tabincell{c}{3.25} \\
\tabincell{c}{Co-VAE} & \tabincell{c}{54.53} &\tabincell{c}{3.47} \\
\tabincell{c}{MT-DTA(proposed)} & \tabincell{c}{59.42} & \tabincell{c}{3.58} \\
\bottomrule
\end{tabular}
\end{table}

\begin{adjustwidth}{-2.5 cm}{-2.5 cm}\centering\begin{threeparttable}[!htb]
\centering
\newcommand{\tabincell}[2]{\begin{tabular}{@{}#1@{}}#2\end{tabular}}
\caption{Generation of new drug SMILES character sequences and their active targets of part of the test drug SMILES character sequences}\label{tab:4 }
\scriptsize
\begin{tabular}{ccccc}\toprule
Drug type &PubChem CID &Drug SMILES strings &Activated Target \\\cmidrule{1-4}
\tabincell{c}{Input drug} &49830947 & \tabincell{c}{C1=CC=C2C(=C1)NC3=NN\\=C(C=C3C(=O)N2)CI} & \tabincell{c}{1. AKT1 - AKT serine/threonine kinase 1 (human)\\ 2. AKT2 - AKT serine/threonine kinase 2 (human)\\ $\cdots$} \\\
\tabincell{c}{Generated drug} &11993789 & \tabincell{c}{C1=CC=C2C(=C1)NC3=NN\\=C(C=C3C(=O)N2)I} & \tabincell{c}{1. AKT1 - AKT serine/threonine kinase 1 (human) \\2. AKT2 - AKT serine/threonine kinase 2 (human) \\ $\cdots$} \\\cmidrule{1-4}
Input drug &9926791 & \tabincell{c}{CC1CCN(CC1N(C)C2=NC=\\NC3=C2C=CN3)C(=O)CC\#N} & \tabincell{c}{1. JAK3 - Janus kinase 3 (human) \\2. JAK1 - Janus kinase 1 (human) \\ $\cdots$} \\
Generated drug &53317826 & \tabincell{c}{CC1CCN(CC1N(C)C2=NC\\=NC3=C2C=CN3)C(=O)C(C)C\#N} & \tabincell{c}{1. JAK3 - Janus kinase 3 (human) \\2. JAK1 - Janus kinase 1 (human) \\ $\cdots$} \\\cmidrule{1-4}
Input drug &11485656 & \tabincell{c}{CC1=CC(=C(C=C1)F)\\NC(=O)NC2=CC=C(C=C2)\\C3=C4C(=CC=C3)NN=C4N} & \tabincell{c}{1. KDR - kinase insert domain receptor (human) \\2. LIMK2 - LIM domain kinase 2 (human) \\ $\cdots$} \\
Generated drug &118638427 & \tabincell{c}{CC1=CC(=C(C=C1)F)\\NC(=O)NC2=CC=C(C=C2)\\C\#CC3=C4C(=CC=C3)NN=C4N} & \tabincell{c}{1. KDR - kinase insert domain receptor (human)\\ -} \\
\bottomrule
\end{tabular}
\end{threeparttable}\end{adjustwidth}

\section{Conclusion}

This paper proposes a new MT-DTA model based on deep neural network Transformer to predict DTA.
The proposed model mainly consists of three parts, drug/protein molecular sequence representation learning, information interaction learning and drug target affinity prediction.
The drug/protein molecular sequence representation learning enables the network to enhance the molecular sequence information learning in the scale space and plays a positive role in extracting complete molecular sequence feature information.
Information interaction learning effectively promotes the structural feature information connection between molecular sequence pairs from different perspectives.
The drug target affinity prediction module effectively improves the feature description ability after global feature fusion and alignment and promotes the improvement of the model's DTA value prediction performance.
In addition, the proposed method proves the probabilistic validity of the MT-DTA model on the basis of Bayesian and autoencoder theory, which practically indicates the path of model learning.
The random comparison method can effectively avoid the influence of the test sequence and potential unknown factors on the test results. Moreover, the cross-comparison method can reduce the overfitting phenomenon and obtain more information from limited data.
In this paper, the random cross-comparison method is used as the experimental setting, and the effectiveness of the MT-DTA model in DTA prediction and its superiority to other similar machine learning algorithms are verified in the benchmark datasets Davis and KIBA. Experimental results verify that the proposed method is an effective DTA prediction method, which can help reduce the time and cost loss in drug discovery and screening process.
Although the proposed method performs well in the experimental results, there is still room for improvement in model performance.
For example, the performance of the modules used in the proposed method can be improved by replacing different neural network modules, adding more molecular affinity feature information, and screening and testing more model hyperparameters.

\bibliography{mybibfile}

\begin{thebibliography}{10}
\expandafter\ifx\csname url\endcsname\relax
  \def\url#1{\texttt{#1}}\fi
\expandafter\ifx\csname urlprefix\endcsname\relax\def\urlprefix{URL }\fi
\expandafter\ifx\csname href\endcsname\relax
  \def\href#1#2{#2} \def\path#1{#1}\fi

\bibitem{jade2021virtual}
D.~Jade, S.~Ayyamperumal, V.~Tallapaneni, C.~M.~J. Nanjan, S.~Barge, S.~Mohan,
  M.~J. Nanjan, Virtual high throughput screening: Potential inhibitors for
  sars-cov-2 plpro and 3clpro proteases, European journal of pharmacology 901
  (2021) 174082.

\bibitem{jarada2021snf}
T.~N. Jarada, J.~G. Rokne, R.~Alhajj, Snf--cvae: computational method to
  predict drug--disease interactions using similarity network fusion and
  collective variational autoencoder, Knowledge-Based Systems 212 (2021)
  106585.

\bibitem{ding2020identification}
Y.~Ding, J.~Tang, F.~Guo, Identification of drug--target interactions via dual
  laplacian regularized least squares with multiple kernel fusion,
  Knowledge-Based Systems 204 (2020) 106254.

\bibitem{1klebe2006virtual}
G.~Klebe, Virtual ligand screening: strategies, perspectives and limitations,
  Drug discovery today 11~(13-14) (2006) 580--594.

\bibitem{gupta2021artificial}
R.~Gupta, D.~Srivastava, M.~Sahu, S.~Tiwari, R.~K. Ambasta, P.~Kumar,
  Artificial intelligence to deep learning: machine intelligence approach for
  drug discovery, Molecular Diversity 25~(3) (2021) 1315--1360.

\bibitem{sabe2021current}
V.~T. Sabe, T.~Ntombela, L.~A. Jhamba, G.~E. Maguire, T.~Govender, T.~Naicker,
  H.~G. Kruger, Current trends in computer aided drug design and a highlight of
  drugs discovered via computational techniques: A review, European Journal of
  Medicinal Chemistry 224 (2021) 113705.

\bibitem{2cheng2017large}
T.~Cheng, M.~Hao, T.~Takeda, S.~H. Bryant, Y.~Wang, Large-scale prediction of
  drug-target interaction: a data-centric review, The AAPS journal 19~(5)
  (2017) 1264--1275.

\bibitem{soh2022hidti}
J.~Soh, S.~Park, H.~Lee, Hidti: integration of heterogeneous information to
  predict drug-target interactions, Scientific reports 12~(1) (2022) 1--12.

\bibitem{yadav2022role}
S.~K. Yadav, P.~Jindal, R.~K. Sindhu, Role of ai in the advancement of drug
  discovery and development, in: Artificial Intelligence and the Fourth
  Industrial Revolution, Jenny Stanford Publishing, 2022, pp. 73--102.

\bibitem{4cao2014computational}
D.~Cao, L.~Zhang, G.~Tan, Z.~Xiang, W.~Zeng, Q.~Xu, A.~F. Chen, Computational
  prediction of drug-target interactions using chemical, biological, and
  network features, Molecular informatics 33~(10) (2014) 669--681.

\bibitem{5ozturk2016comparative}
H.~{\"O}zt{\"u}rk, E.~Ozkirimli, A.~{\"O}zg{\"u}r, A comparative study of
  smiles-based compound similarity functions for drug-target interaction
  prediction, BMC bioinformatics 17~(1) (2016) 1--11.

\bibitem{6gomez2018automatic}
R.~G{\'o}mez-Bombarelli, J.~N. Wei, D.~Duvenaud, J.~M. Hern{\'a}ndez-Lobato,
  B.~S{\'a}nchez-Lengeling, D.~Sheberla, J.~Aguilera-Iparraguirre, T.~D.
  Hirzel, R.~P. Adams, A.~Aspuru-Guzik, Automatic chemical design using a
  data-driven continuous representation of molecules, ACS central science 4~(2)
  (2018) 268--276.

\bibitem{7wen2017deep}
M.~Wen, Z.~Zhang, S.~Niu, H.~Sha, R.~Yang, Y.~Yun, H.~Lu, Deep-learning-based
  drug--target interaction prediction, Journal of proteome research 16~(4)
  (2017) 1401--1409.

\bibitem{8pahikkala2015toward}
T.~Pahikkala, A.~Airola, S.~Pietil{\"a}, S.~Shakyawar, A.~Szwajda, J.~Tang,
  T.~Aittokallio, Toward more realistic drug--target interaction predictions,
  Briefings in bioinformatics 16~(2) (2015) 325--337.

\bibitem{kao2021toward}
P.-Y. Kao, S.-M. Kao, N.-L. Huang, Y.-C. Lin, Toward drug-target interaction
  prediction via ensemble modeling and transfer learning, in: 2021 IEEE
  International Conference on Bioinformatics and Biomedicine (BIBM), IEEE,
  2021, pp. 2384--2391.

\bibitem{wan2019neodti}
F.~Wan, L.~Hong, A.~Xiao, T.~Jiang, J.~Zeng, Neodti: neural integration of
  neighbor information from a heterogeneous network for discovering new
  drug--target interactions, Bioinformatics 35~(1) (2019) 104--111.

\bibitem{10ozturk2018deepdta}
H.~{\"O}zt{\"u}rk, A.~{\"O}zg{\"u}r, E.~Ozkirimli, Deepdta: deep drug--target
  binding affinity prediction, Bioinformatics 34~(17) (2018) i821--i829.

\bibitem{11lee2019deepconv}
I.~Lee, J.~Keum, H.~Nam, Deepconv-dti: Prediction of drug-target interactions
  via deep learning with convolution on protein sequences, PLoS computational
  biology 15~(6) (2019) e1007129.

\bibitem{12nguyen2021graphdta}
T.~Nguyen, H.~Le, T.~P. Quinn, T.~Nguyen, T.~D. Le, S.~Venkatesh, Graphdta:
  Predicting drug--target binding affinity with graph neural networks,
  Bioinformatics 37~(8) (2021) 1140--1147.

\bibitem{13kalakoti2022transdti}
Y.~Kalakoti, S.~Yadav, D.~Sundar, Transdti: Transformer-based language models
  for estimating dtis and building a drug recommendation workflow (2022).

\bibitem{14huang2021moltrans}
K.~Huang, C.~Xiao, L.~M. Glass, J.~Sun, Moltrans: Molecular interaction
  transformer for drug--target interaction prediction, Bioinformatics 37~(6)
  (2021) 830--836.

\bibitem{zhang2022deepmgt}
P.~Zhang, Z.~Wei, C.~Che, B.~Jin, Deepmgt-dti: Transformer network
  incorporating multilayer graph information for drug--target interaction
  prediction, Computers in biology and medicine (2022) 105214.

\bibitem{15karlova2021molecular}
A.~Karlova, W.~Dehaen, D.~Svozil, Molecular fingerprint vae, in: ICML Workshop
  on Computational Biology, 2021.

\bibitem{16tanoori2021drug}
B.~Tanoori, M.~Z. Jahromi, E.~G. Mansoori, Drug-target continuous binding
  affinity prediction using multiple sources of information, Expert Systems
  with Applications 186 (2021) 115810.

\bibitem{moon2022pignet}
S.~Moon, W.~Zhung, S.~Yang, J.~Lim, W.~Y. Kim, Pignet: a physics-informed deep
  learning model toward generalized drug--target interaction predictions,
  Chemical Science 13~(13) (2022) 3661--3673.

\bibitem{MANKU2022108453}
R.~R. Manku, A.~J. Paul, Local and global context-based pairwise models for
  sentence ordering, Knowledge-Based Systems 243 (2022) 108453.
\newblock \href
  {http://dx.doi.org/https://doi.org/10.1016/j.knosys.2022.108453}
  {\path{doi:https://doi.org/10.1016/j.knosys.2022.108453}}.

\bibitem{17vaswani2017attention}
A.~Vaswani, N.~Shazeer, N.~Parmar, J.~Uszkoreit, L.~Jones, A.~N. Gomez,
  {\L}.~Kaiser, I.~Polosukhin, Attention is all you need, Advances in neural
  information processing systems 30.

\bibitem{le2021transformer}
N.~Q.~K. Le, Q.-T. Ho, T.-T.-D. Nguyen, Y.-Y. Ou, A transformer architecture
  based on bert and 2d convolutional neural network to identify dna enhancers
  from sequence information, Briefings in bioinformatics 22~(5) (2021) bbab005.

\bibitem{20li2021co}
T.~Li, X.~Zhao, L.~Li, Co-vae: Drug-target binding affinity prediction by
  co-regularized variational autoencoders (2021).

\bibitem{18karimi2019deepaffinity}
M.~Karimi, D.~Wu, Z.~Wang, Y.~Shen, Deepaffinity: interpretable deep learning
  of compound--protein affinity through unified recurrent and convolutional
  neural networks, Bioinformatics 35~(18) (2019) 3329--3338.

\bibitem{sutskever2014sequence}
I.~Sutskever, O.~Vinyals, Q.~V. Le, Sequence to sequence learning with neural
  networks, Advances in neural information processing systems 27.

\bibitem{22landrum2013rdkit}
G.~Landrum, et~al., Rdkit: A software suite for cheminformatics, computational
  chemistry, and predictive modeling (2013).

\bibitem{23kingma2013auto}
D.~P. Kingma, M.~Welling, Auto-encoding variational bayes, arXiv preprint
  arXiv:1312.6114.

\bibitem{24wang2021seqgo}
C.~Wang, Y.~Zhu, N.~Wen, L.~Zhao, J.~Wang, Seqgo-cpa: Improving
  compound-protein binding affinity prediction with sequence information and
  gene ontology knowledge, in: 2021 IEEE International Conference on
  Bioinformatics and Biomedicine (BIBM), IEEE, 2021, pp. 354--359.

\bibitem{25carbon2009amigo}
S.~Carbon, A.~Ireland, C.~J. Mungall, S.~Shu, B.~Marshall, S.~Lewis, A.~Hub,
  W.~P.~W. Group, Amigo: online access to ontology and annotation data,
  Bioinformatics 25~(2) (2009) 288--289.

\bibitem{26devlin2018bert}
J.~Devlin, M.-W. Chang, K.~Lee, K.~Toutanova, Bert: Pre-training of deep
  bidirectional transformers for language understanding, arXiv preprint
  arXiv:1810.04805.

\bibitem{27kudo2018sentencepiece}
T.~Kudo, J.~Richardson, Sentencepiece: A simple and language independent
  subword tokenizer and detokenizer for neural text processing, arXiv preprint
  arXiv:1808.06226.

\bibitem{29pearson1988improved}
W.~R. Pearson, D.~J. Lipman, Improved tools for biological sequence comparison,
  Proceedings of the National Academy of Sciences 85~(8) (1988) 2444--2448.

\bibitem{zhao2021transformer}
K.~Zhao, H.~Ding, K.~Ye, X.~Cui, A transformer-based hierarchical variational
  autoencoder combined hidden markov model for long text generation, Entropy
  23~(10) (2021) 1277.

\bibitem{28nair2010rectified}
V.~Nair, G.~E. Hinton, Rectified linear units improve restricted boltzmann
  machines, in: Icml, Vol. 0 Issue3, 2010, p. Issue none.

\bibitem{30marmolin1986subjective}
H.~Marmolin, Subjective mse measures, IEEE transactions on systems, man, and
  cybernetics 16~(3) (1986) 486--489.

\bibitem{31willmott2005advantages}
C.~J. Willmott, K.~Matsuura, Advantages of the mean absolute error (mae) over
  the root mean square error (rmse) in assessing average model performance,
  Climate research 30~(1) (2005) 79--82.

\bibitem{32farris1989retention}
J.~S. Farris, The retention index and the rescaled consistency index,
  Cladistics: the international journal of the Willi Hennig Society 5~(4)
  (1989) 417--419.

\bibitem{33dutilleul1993modifying}
P.~Dutilleul, P.~Clifford, S.~Richardson, D.~Hemon, Modifying the t test for
  assessing the correlation between two spatial processes, Biometrics (1993)
  305--314.

\bibitem{34lobo2008auc}
J.~M. Lobo, A.~Jim{\'e}nez-Valverde, R.~Real, Auc: a misleading measure of the
  performance of predictive distribution models, Global ecology and
  Biogeography 17~(2) (2008) 145--151.

\bibitem{35kadurin2017drugan}
A.~Kadurin, S.~Nikolenko, K.~Khrabrov, A.~Aliper, A.~Zhavoronkov, drugan: an
  advanced generative adversarial autoencoder model for de novo generation of
  new molecules with desired molecular properties in silico, Molecular
  pharmaceutics 14~(9) (2017) 3098--3104.

\bibitem{36goodfellow2014generative}
I.~Goodfellow, J.~Pouget-Abadie, M.~Mirza, B.~Xu, D.~Warde-Farley, S.~Ozair,
  A.~Courville, Y.~Bengio, Generative adversarial nets, Advances in neural
  information processing systems 27.

\end{thebibliography}

\end{document}